\documentclass[
reprint, 
superscriptaddress,
 amsmath,amssymb,
  aps,
pre,
]{revtex4-1}

\usepackage{graphicx}
\usepackage{dcolumn}
\usepackage{bm}
\usepackage{todonotes}

\usepackage{color}

\newcommand{\diff}[2]{\frac{d#1}{d#2}}
\newcommand{\pdiff}[2]{\frac{\partial #1}{\partial #2}}

\newcommand{\new}{\nonumber\\}
\newcommand{\abs}[1]{\left|#1\right|}

\newcommand{\bu}{\bm{u}}
\newcommand{\bn}{\bm{n}}
\newcommand{\br}{\bm{r}}

\newcommand{\HH}{\mathcal{H}}

\usepackage{amsmath}	
\begin{document}

\preprint{APS/123-Qed} \title{Testing mean-field theory for jamming of
non-spherical particles: Contact number, gap distribution, and
vibrational density of states}


\author{Harukuni Ikeda}
 \email{hikeda@g.ecc.u-tokyo.ac.jp}
\affiliation{Graduate School of Arts and Sciences, The University of
Tokyo 153-8902, Japan}


\date{\today}
	     
\begin{abstract} 
We perform numerical simulations of the jamming transition of
non-spherical particles in two dimensions. In particular, we
systematically investigate how the physical quantities at the jamming
transition point behave when the shapes of the particle deviate slightly
from the perfect disks. For efficient numerical simulation, we first
derive an analytical expression of the gap function, using the
perturbation theory around the reference disks. Starting from disks, we
observe the effects of the deformation of the shapes of particles by the
$n$-th order term of the Fourier series $\sin(n\theta)$. We show that
the several physical quantities, such as the number of contacts, gap
distribution, and characteristic frequencies of the vibrational density
of states, show the power-law behaviors with respect to the linear
deviation from the reference disks. The power-law behaviors do not
depend on $n$ and are fully consistent with the mean-field theory of the
jamming of non-spherical particles. This result suggests that the
mean-field theory holds very generally for nearly spherical particles
whose shape can be expressed by the Fourier series.
\end{abstract}

\maketitle

\section{Introduction}

The jamming transition is a phenomenon that granular materials suddenly
get finite rigidity at a certain density referred to as the jamming
transition point $\varphi_J$~\cite{van2009jamming,liu2010jamming}. Near
$\varphi_J$, several physical quantities, such as energy, mechanical
pressure, shear modulus, and contact number, exhibit the power-law
behaviors~\cite{ohern2003}. This implies that the jamming transition is a
critical phenomenon~\cite{van2009jamming,liu2010jamming}, such as the
second-order phase transition of equilibrium
systems~\cite{nishimori2010elements}.

The simplest model to study the jamming transition is the system
consisting of frictionless spherical particles. The systematic numerical
simulations confirmed that the critical exponents of several physical
quantities do not depend on the spatial dimensions $d$ for $d\geq
2$~\cite{ohern2003,vaagberg2011finite,charbonneau2014fractal}, while
different exponents appear in a quasi-one-dimensional
system~\cite{hikeda2020}. The dimensional independence of the critical
exponents suggests that the exact values of the critical exponents can
be calculated by considering the large dimensional limit $d\to\infty$,
where the mean-field theory becomes
exact~\cite{charbonneau2014fractal,parisi2020theory}. This calculation
has been done by using the replica method, which was originally
developed for the spin-glasses, but is now widely used for many disordered
systems~\cite{mezard1987spin,nishimori2001statistical}. Resultant
critical exponents well agree with the numerical results in $d\geq
2$~\cite{charbonneau2014fractal,parisi2020theory}. Several other
mean-field theories, such as the variational
argument~\cite{wyart2005effects,yan2016variational}, effective medium
theory~\cite{degiuli2014effects,degiuli2014force}, and random matrix
theory~\cite{beltukov2015random,ikeda2020note}, are also successfully
applied to derive the scaling laws of the jamming transition.

Motivated by the success of the mean-field theories for the jamming of
spherical particles, we recently developed a mean-field theory of the
jamming of nearly spherical
particles~\cite{brito2018universality,ikeda2019mean}. The theory
predicts that several physical quantities, such as the contact number
and gap distribution, exhibit the singular behaviors in the limit of
perfect spheres. The theoretical conjectures, however, have been
confirmed mainly for the \textit{breathing} particles, which are
believed to have the same universality class of non-spherical
particles~\cite{brito2018universality}. It is, of course, desirable to
perform a more direct test for  non-spherical particles.
That is the purpose of this work.

The problem is that for particles of general shape, it is numerically
demanding and technically involved to calculate the gap function, which
is the minimal distance between two particles and used to judge if two
particles are overlapped or not~\cite{lu2015discrete}. However, since we
are interested in the case where the particle shapes are close to
perfect spheres, we can derive an analytic form of the gap function by
using a perturbation expansion from the reference
spheres~\cite{ikeda2020infinitesimal,tunable2020}. This allows us to
perform an efficient numerical simulation.

We consider a particle system in two dimensions $d=2$. Starting
from perfect disks, we deform the shapes of particles by the $n$-th
order term of the Fourier series $\propto \sin(n\theta)$. We observe how
this deformation affects the physical quantities, such as the contact
number, gap distribution function, and characteristic frequencies of the
vibrational density of states, at $\varphi_J$. We find that the
qualitative behaviors do not depend on $n$ and fully consistent with the
mean-field predictions.

The organization of the paper is as follows. In
Sec.~\ref{132018_25Nov20}, we summarize the previous results for jamming
of frictionless spherical and non-spherical particles. In
Sec.~\ref{132227_25Nov20}, we introduce the model and several important
physical quantities. In Sec.~\ref{132722_25Nov20}, we discuss the
approximation to calculate the gap function and interaction
potential. In Sec.~\ref{132255_25Nov20}, we discuss the numerical
algorithm to generate configurations at the jamming transition point. In
Sec.~\ref{132327_25Nov20}, we check the validity of our approximation by
comparing the result of our model for $n=2$ with a previous result of
ellipses. In Sec.~\ref{133030_25Nov20}, we discuss the behavior of
$\varphi_J$ and the fraction of rattles. In Sec.~\ref{133113_25Nov20}, we
discuss the scaling of the contact number at the jamming transition
point. In Sec.~\ref{133141_25Nov20}, we discuss the scaling of the gap
function. In Sec.~\ref{133202_25Nov20}, we discuss the behavior of the
vibrational density of states. Finally, in Sec.~\ref{133229_25Nov20}, we
summarize the results and conclude the work.

\section{Previous results}
\label{132018_25Nov20}

We here summarize the previous numerical and theoretical results for the
jamming transition of frictionless spherical and non-spherical
particles.

\subsection{Spherical particles}

Frictionless spherical particles are \textit{isostatic} at the jamming
transition point in the thermodynamics limit. We first explain the
concept of the isostaticity. A system is said to be isostatic if it
satisfies the following condition:
\begin{align}
N_c = N_f,\label{135028_18Nov20}
\end{align}
where $N_c$ denotes the total number of constraints of the system, and
$N_f$ denotes the total number of degrees of
freedom. For $N$ frictionless
spherical particles in $d$ dimensions, the total number of
degrees of freedom is
\begin{align}
 N_f = Nd.\label{135115_18Nov20}
\end{align}
For a jammed configuration generated by the isotropic compression, 
$N_c$ is written as 
\begin{align}
 N_c = \frac{Nz}{2} + d,\label{090611_23Nov20}
\end{align}
where $z$ denotes the number of contacts per particle.  In
Eq.~(\ref{090611_23Nov20}), the first term $Nz/2$ represents the total
number of contacts, and the second term $d$ comes from the requirement
for the positive bulk modulus~\cite{correction2011}. Therefore, if the
system is isostatic, the contact number per particles is
\begin{align}
 z_{\rm iso} = 2d-\frac{2d}{N}.
\end{align}
The systematic numerical simulations revealed that frictionless
spherical particles are almost isostatic at $\varphi_J$, more precisely,
the contact number at $\varphi_J$, $z_J\equiv z(\varphi_J)$, is~\cite{goodrich2012}
\begin{align}
 z_J-z_{\rm iso}\propto \frac{1}{N}.
\end{align}
In particular, $z_J\to z_{\rm
iso}$ in the thermodynamic limit $N\to\infty$~\cite{bernal1960packing,ohern2003}.

$z$ increases on increasing the packing fraction $\varphi$. Numerical
studies near $\varphi_J$ revealed that $z$ exhibits the following
power-law behavior~\cite{ohern2003,goodrich2012}:
\begin{align}
 \delta z \equiv z-z_{\rm iso} \propto \delta\varphi^{1/2},\label{151508_18Nov20}
\end{align}
where $\delta\varphi = \varphi-\varphi_J$. Another interesting
power-law appears if one observes the gap distribution,
\begin{align}
 g(h) = \frac{1}{N}\sum_{i<j}\Theta(-h_{ij})\delta(h-h_{ij}),
\end{align}
where $\Theta(x)$ denotes the Heaviside step function, and $h_{ij}$
denotes the gap function defined as
\begin{align}
 h_{ij} = \abs{\br_i-\br_j}-R_i-R_j.
\end{align}
Here $\br_i$ and $R_i$ denote the position and radius of the $i$-th
particle, respectively. At $\varphi_J$ in the thermodynamic limit,
$g(h)$ exhibits the power law for $h\ll
1$~\cite{donev2005,charbonneau2012}:
\begin{align}
 g(h) \sim h^{-\gamma}, 
\end{align}
where $\gamma \approx
0.41$~\cite{charbonneau2014fractal,charbonneau2020finite}. For
$\varphi>\varphi_J$, on the contrary, the power-law is truncated at
finite $h$~\cite{charbonneau2012,franz2017universality}:
\begin{align}
 g(h)
\sim
\begin{cases}
\delta\varphi^{-\mu\gamma} & h\ll \delta\varphi^{\mu} \\
h^{-\gamma} & h\gg \delta\varphi^{\mu}
\end{cases},
\end{align}
where
\begin{align}
 \mu = \frac{1}{2(1-\gamma)}.
\end{align}
\begin{figure}[t]
\begin{center}
 \includegraphics[width=8cm]{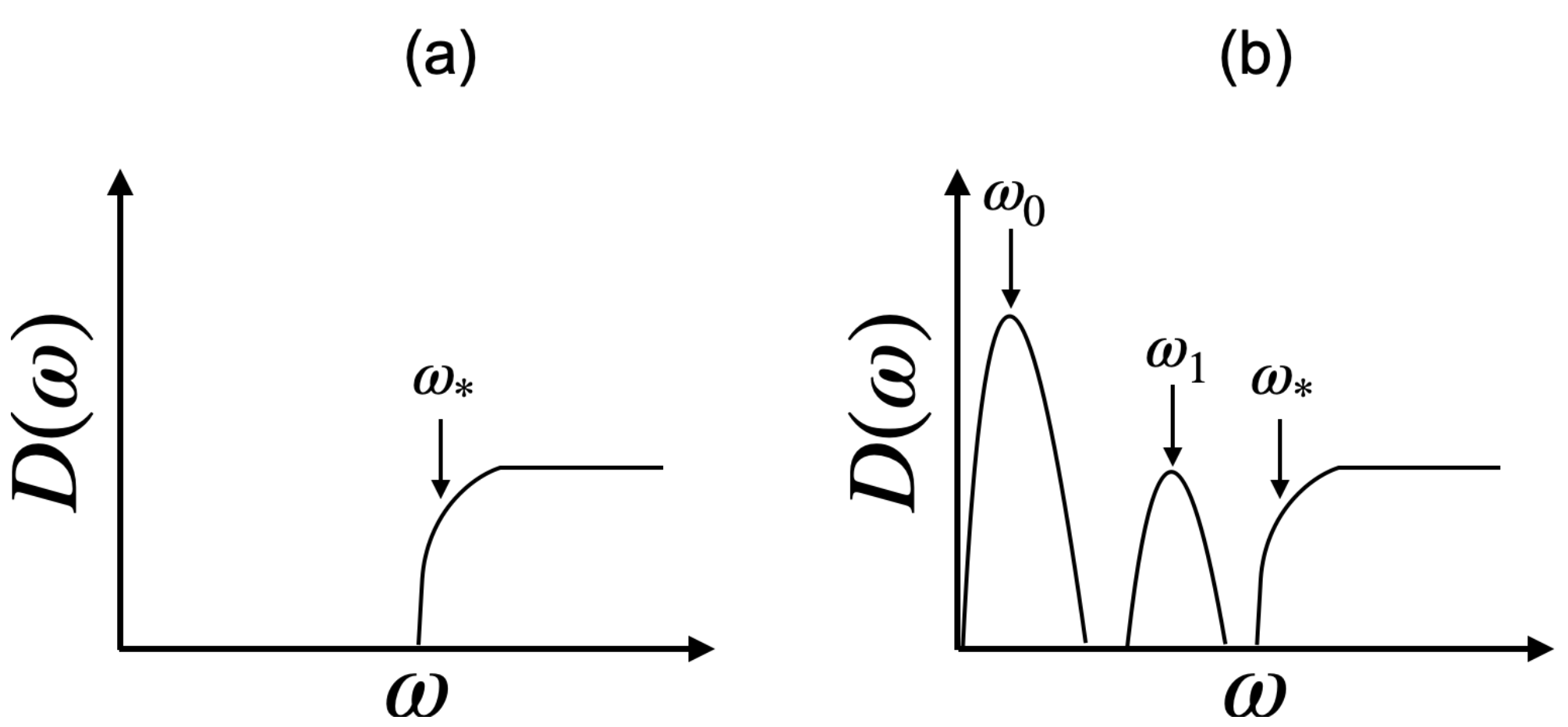} \caption{ Schematic picture
 of $D(\omega)$. (a) $D(\omega)$ of spherical particles. (b) $D(\omega)$
 of non-spherical particles.}
					    \label{120311_23Nov20}
\end{center}
\end{figure}
Finally, in Fig.~\ref{120311_23Nov20}~(a), we show the schematic figure
of the vibrational density of states near $\varphi_J$. $D(\omega)$
exhibits a plateau down to the characteristic frequency
$\omega_*$. $\omega_*$ exhibits the following power-law
behavior~\cite{ohern2003,wyart2005effects}:
\begin{align}
 \omega_* \sim \delta z \sim \delta\varphi^{1/2}.\label{174734_25Nov20}
\end{align}

Several mean-field theories, such as the variational
argument~\cite{yan2016variational,wyart2005effects}, effective medium
theory~\cite{degiuli2014effects,degiuli2014force}, and replica
method~\cite{charbonneau2014fractal,franz2017universality}, have been
successful in reproducing the above scaling behaviors of the jamming of
frictionless spherical particles. This motivates us to develop the
mean-field theories for the jamming of non-spherical particles, which we
shall discuss in the next subsection.

\subsection{non-spherical particles}

The vibrational argument predicts the same scaling laws as
Eqs.~(\ref{151508_18Nov20}) and (\ref{174734_25Nov20}), for a system
satisfying the isostaticity Eq.~(\ref{135028_18Nov20}) at
$\varphi_J$~\cite{yan2016variational}. The same argument holds even for
non-spherical particles. For instance, frictionless dimers become
isostatic at $\varphi_J$~\cite{schreck2010comparison}, and a systematic
numerical simulation indeed confirmed the same scaling laws as spherical
particles~\cite{shiraishi2019,shiraishi2020mechanical}. More generally,
for non-spherical particles consisting of spherical particles, such as
dimers, trimers, \dots, $n$-mers, one can show that the systems are
isostatic at $\varphi_J$ (see Appendix.~\ref{075105_23Nov20} for a
theoretical argument).

On the contrary, the deformation from the perfect sphere (or dimer,
trimer, \dots), in general, breaks the isostaticity. For instance,
frictionless
ellipsoids~\cite{donev2004improving,donev2007underconstrained,zeravcic2009excitations,mailman2009jamming,schreck2010comparison,schreck2012constraints},
spherocylinders~\cite{williams2003random,blouwolff2006coordination,stress2010,marschall2018compression},
superballs~\cite{jiao2010}, superellipsoids~\cite{delaney2010packing},
and circulo-polygons~\cite{vander2018} have been known to be
\textit{hypostatic} $N_c<N_f$ at $\varphi_J$. In the previous works, we
have constructed the mean-field theories and derived the scaling laws
for non-spherical particles, which are hypostatic at
$\varphi_J$~\cite{brito2018universality,ikeda2019mean,ikeda2020infinitesimal}.
Below we summarize the main results.

The diameter $R$ of a non-spherical particle depends on the direction
$\bn$. We shall write $R$ as
\begin{align}
 R(\bn,\Delta) = R_0\left(1 + \Delta F(\bn)\right),
\end{align}
where $R_0$ denotes the radius of the reference sphere, $\Delta$
represents the linear deviation from the reference sphere, and $F(\bn)$
characterizes the particle shape. The mean-field theory predicts that
for $\Delta\ll 1$, the contact number at $\varphi_J$ behaves as
\begin{align}
 \delta z_J = z_J(\Delta)-z_{\rm iso} \propto \Delta^{1/2},\label{152428_18Nov20}
\end{align}
and $g(h)$ behaves as
\begin{align}
 g(h) \sim
\begin{cases}
\Delta^{-\mu\gamma} & h\ll \Delta^{\mu} \\
h^{-\gamma} & h\gg \Delta^{\mu}
\end{cases}.\label{152432_18Nov20}
\end{align}
In Fig.~\ref{120311_23Nov20}, we also show the schematic
picture of $D(\omega)$ of non-spherical particles. As in the case of
spherical particles, $D(\omega)$ exhibits a plateau down to
$\omega_*$. In addition, $D(\omega)$ of non-spherical particles has two
additional peaks at $\omega_0, \omega_1\ll \omega_*$. The
characteristic frequencies exhibit the power-law behaviors:
\begin{align}
\omega_0 \sim \Delta^{1/2}\delta\varphi^{1/2},\
\omega_1 \sim \Delta,\
 \omega_* \sim \Delta^{1/2}.\label{123636_23Nov20}
\end{align}
In particular, at $\varphi_J$, we get 
\begin{align}
  \omega_0  = 0,\
  \omega_1 \sim \Delta,\
  \omega_* \sim \Delta^{1/2}.\label{122941_23Nov20}
\end{align}
This suggests that the modes in the lowest band become zero modes at
$\varphi_J$. These zero modes are consequence of the hypostaticity
$N_c<N_f$ and referred to as the quartic
modes~\cite{mailman2009jamming}. Eqs.~(\ref{152428_18Nov20}),
(\ref{152432_18Nov20}), and (\ref{122941_23Nov20}) imply that increasing
$\Delta$ at $\varphi_J$ causes the same results as increasing
$\delta\varphi$ of spherical particles.

So far, Eq.~(\ref{152428_18Nov20}) has been checked only for systems
with small system size $N=100$, where the power-law is barely
visible~\cite{brito2018universality}, Eq.~(\ref{152432_18Nov20}) has not
been checked for non-spherical particles, and Eq.~(\ref{122941_23Nov20})
has been checked only for ellipsoids~\cite{zeravcic2009excitations}.
The purpose of this work is to test the mean-field predictions for
general shapes of non-spherical particles at $\varphi_J$.

\section{Settings}
\label{132227_25Nov20} Here we introduce the model and several important
physical quantities.

\subsection{Model} 
\begin{figure*}[t]
\begin{center}
 \includegraphics[width=18cm]{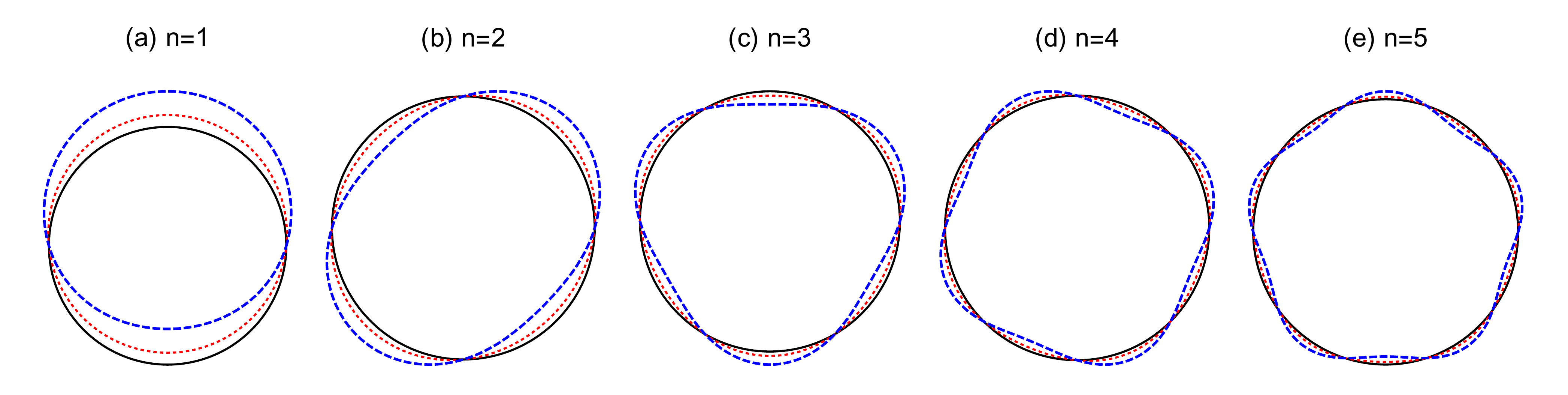} \caption{Particle shapes for
 various $n$ and $\Delta$. Black solid, red dotted, and blue dashed
 lines denote polar plots of $R_i(\theta)/R_i^0=1+\Delta F_n(\theta)$
 for $\Delta=0$, $0.1$, and $0.3$, respectively. }
 \label{034905_14Nov20}
\end{center}
\end{figure*}

We consider a two dimensional system consisting of $N$ non-spherical
particles. The radius of a non-spherical particle depends on its angle
$\theta\in [0,2\pi)$. We shall write the radius of the $i$-th particle
as
\begin{align}
 R_i(\theta) = R_i^0\left[1+ \Delta F(\theta)\right],\label{131104_23Nov20}
\end{align}
where $R_i^0$ denotes the radius of the reference disk defined by
\begin{align}
 R_i^0 = \frac{1}{2\pi}\int_0^{2\pi}d\theta R_i(\theta).\label{131111_23Nov20}
\end{align}
$\Delta$ represents the linear deviation from the reference disk, and
the function $F(\theta)$ characterizes the particle shape. $F(\theta)$
satisfies
\begin{align}
 \int_0^{2\pi}d\theta F(\theta) = 0.\label{130422_23Nov20}
\end{align}
Since $F(\theta)$ is a periodic function, it is natural to express
$F(\theta)$ by the Fourier series as
\begin{align}
 F(\theta) = \sum_{n=1}^\infty a_n\sin (n\theta + b_n),
\end{align}
where the constant term does not appear due to
Eq.~(\ref{130422_23Nov20}). Starting from the reference disk
$R_i(\theta)=R_i^0$, we want to investigate how the $n$-th term of the
Fourier series perturbs the physical quantities at $\varphi_J$. For
this purpose, we
shall consider the following functional form of $R_i(\theta)$:
\begin{align}
&R_i(\theta) = R_i^0\left[1+\Delta F_n(\theta)\right],\label{133613_23Nov20}\\
&F_n(\theta)  = \frac{\sin(n\theta)}{n}.
\end{align}
When $\Delta=0$, the $i$-th particle is, of course, a disk of the radius
$R_i(\theta)=R_i^0$. On increasing $\Delta$, the shape of the $i$-th
particle gradually deviates from the reference disk. In
Fig.~\ref{034905_14Nov20}, we illustrate the corresponding particle
shape for various $n$ and $\Delta$.

\subsection{Volume fraction}

We define the volume fraction $\varphi$ as
\begin{align}
 \varphi = \frac{\sum_{i=1}^N S_i}{L^2},
\end{align}
where $L$ denotes the linear distance of the system, and $S_i$ denotes
the surface of the $i$-th particle calculated as
\begin{align}
S_i = \int_0^{2\pi}d\theta \int_0^{R_i(\theta)}r dr
 = \frac{(R_i^0)^2}{2} \int_0^{2\pi}d\theta
 \left[1+\Delta F_n(\theta)\right]^2.
\end{align}

\subsection{Asphericity}
For a particle of general shape, it is not always straightforward to
find out the parameter corresponding to the linear deviation from the
reference disk $\Delta$. It is sometimes convenient to use the
asphericity:
\begin{align}
 A_i =  \frac{P_i^2}{4\pi S_i},\label{133620_23Nov20}
\end{align}
where $P_i$ denotes the perimeter calculated as 
\begin{align}
 P_i = \int_0^{2\pi}d\theta \sqrt{R_i(\theta)^2 
+ \left(\diff{R_i(\theta)}{\theta}\right)^2}d\theta.
\end{align}
$A_i$ is calculated straightforwardly for particles of any shape.
Furthermore, it is possible to derive the scaling relation between $A_i$
and $\Delta$, as follows. $A_i$ takes the minimal value $A_i=1$ for a
disk $\Delta=0$, and $A_i>1$ for a non-disk $\Delta\neq 0$. Therefore,
we get
\begin{align}
& A_i(\Delta) = 1 + \frac{\Delta^2}{2} A_i''(\Delta) \\
&\to \Delta  \propto (A_i-1)^{1/2},\label{152315_10Dec20}
\end{align}
where the linear order term does not appear, as $A_i(\Delta)$ has a
minimum at $\Delta=0$.  Eq.~(\ref{152315_10Dec20}) allows to convert the
scaling laws for $\Delta$
Eqs.~(\ref{152428_18Nov20},\ref{152432_18Nov20},\ref{123636_23Nov20}) to
that for $A_i$~\cite{ikeda2020infinitesimal}. By substituting
Eq.~(\ref{133613_23Nov20}) into Eq.~(\ref{133620_23Nov20}), one can see
that $A_i$ does not depend on $i$, $A_i=A$. For $\Delta\ll 1$, we get
\begin{align}
A &\approx 1 + \frac{\Delta^2}{2\pi}\int d\theta\left(F_n'(\theta)^2-F_n(\theta)^2\right)\\
 &= 1 + \frac{\Delta^2}{2}\left(1-\frac{1}{n^2}\right).\label{061117_15Nov20}
\end{align}
Note that $O(\Delta^2)$ order term vanishes for $n=1$. This means that
the $n=1$ Fourier component only causes the translation of a particle
for the lowest order correction~\cite{tarama2013dynamics}, as
illustrated in Fig.~\ref{034905_14Nov20}~(a). Hereafter we only consider
$n>1$. Later, we use Eq.~(\ref{061117_15Nov20}) to compare our result
with the previous work of ellipsoids.

\section{Interaction potential}
\label{132722_25Nov20}

\begin{figure}[t]
\begin{center}
 \includegraphics[width=8cm]{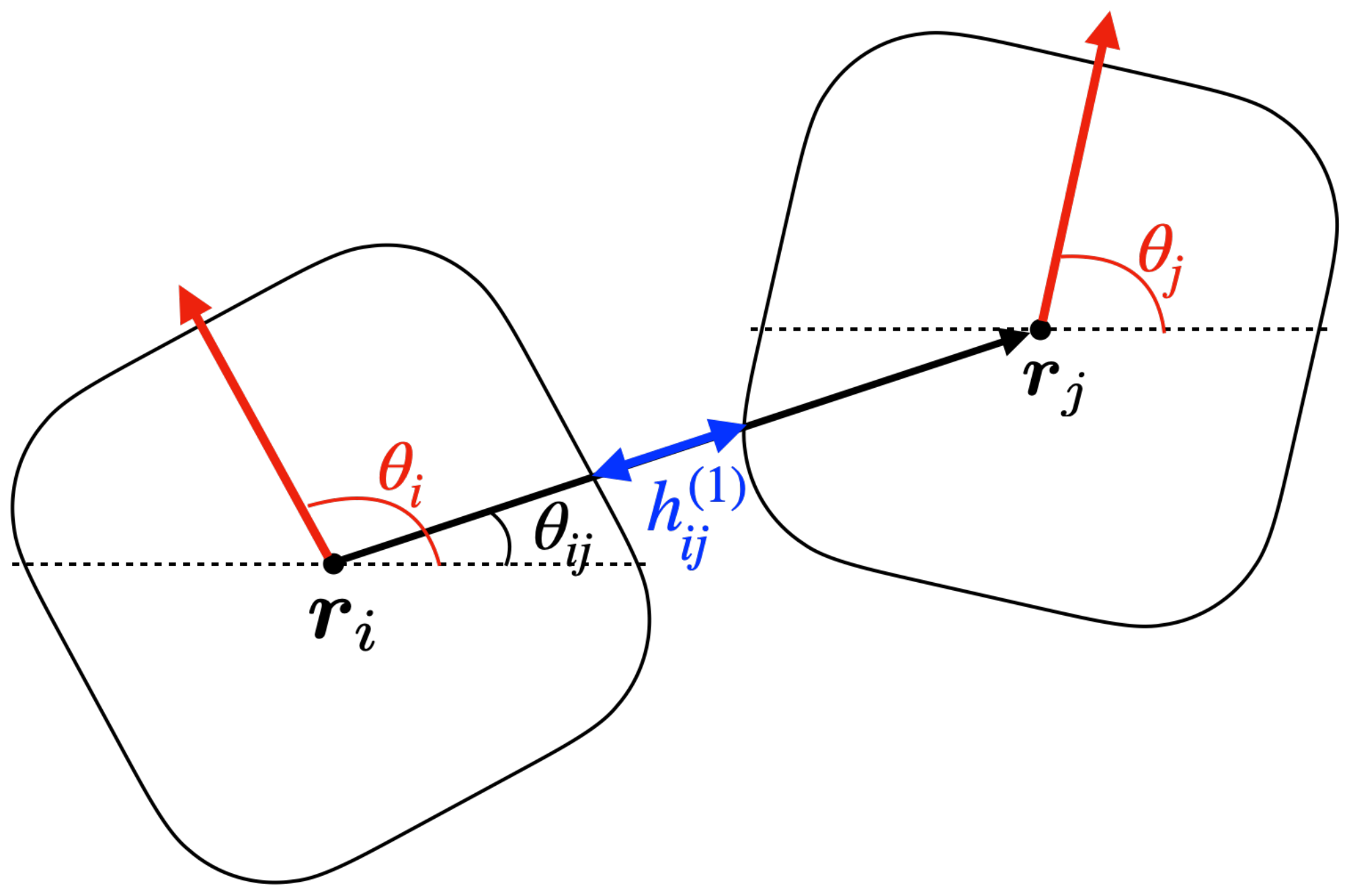} \caption{Definitions of
 $\br_i$, $\br_j$, $\theta_i$, $\theta_j$, $\theta_{ij}$, and
 $h_{ij}^{(1)}$. Red arrows denote particles directions, black arrow
 denotes the vector $\br_j-\br_i$, and dashed lines are parallel to the
 $x$-axis.}  \label{173935_24Nov20}
\end{center}
\end{figure}

We consider the harmonic potential~\cite{ohern2003}:
\begin{align}
 V_N = \sum_{i<j}\frac{h_{ij}^2}{2}\Theta(-h_{ij}),
\end{align}
where $\Theta(x)$ denotes the Heaviside step function, and $h_{ij}$
denotes the gap function, which is the minimal distance between
particles $i$ and $j$. In general, it is a non-trivial task to calculate
$h_{ij}$ for general shapes of non-spherical particles. Since we are
interested in the scaling behaviors for $\Delta \ll 1$, we calculate
$h_{ij}$ by using the first order expansion w.r.t $\Delta$, see
Appendix.~\ref{131917_23Nov20} for details of the calculation. After
some manipulations, we get
\begin{align}
 h_{ij} &= 
 h_{ij}^{(1)} + O(\Delta^2),\\
 h_{ij}^{(1)} &=
\abs{\br_i-\br_j}-R_i^0 -R_j^0\new
&-\Delta\left[R_i^0 F_n(\theta_i-\theta_{ij})
 +R_j^0 F_n(\theta_j-\theta_{ji})\right],\label{084534_15Nov20}
\end{align}
where
$\br_i=\{x_i,y_i\}$ denotes the position of the $i$-th particle, 
$\theta_i$ denotes the direction of the $i$-th particle, and
$\theta_{ij}$ denotes the relative angle of particles $i$ and $j$,
namely, $\theta_{ij}={\rm atan2}(y_j-y_i,x_j-x_i)$, 
see Fig.~\ref{173935_24Nov20}.
Hereafter, we use Eq.~(\ref{084534_15Nov20}) to calculate the
interaction potential. Obviously, the approximation holds only for
$\Delta \ll 1$. Another limitation is that by construction, our
approximation allows that two particles have at most a single contact,
which is not true for particles of non-convex shapes, such as dimers,
even for small aspect ratios~\cite{schreck2010comparison}.

\section{Numerics}
\label{132255_25Nov20} 
\begin{figure}[t]
\begin{center}
 \includegraphics[width=8cm]{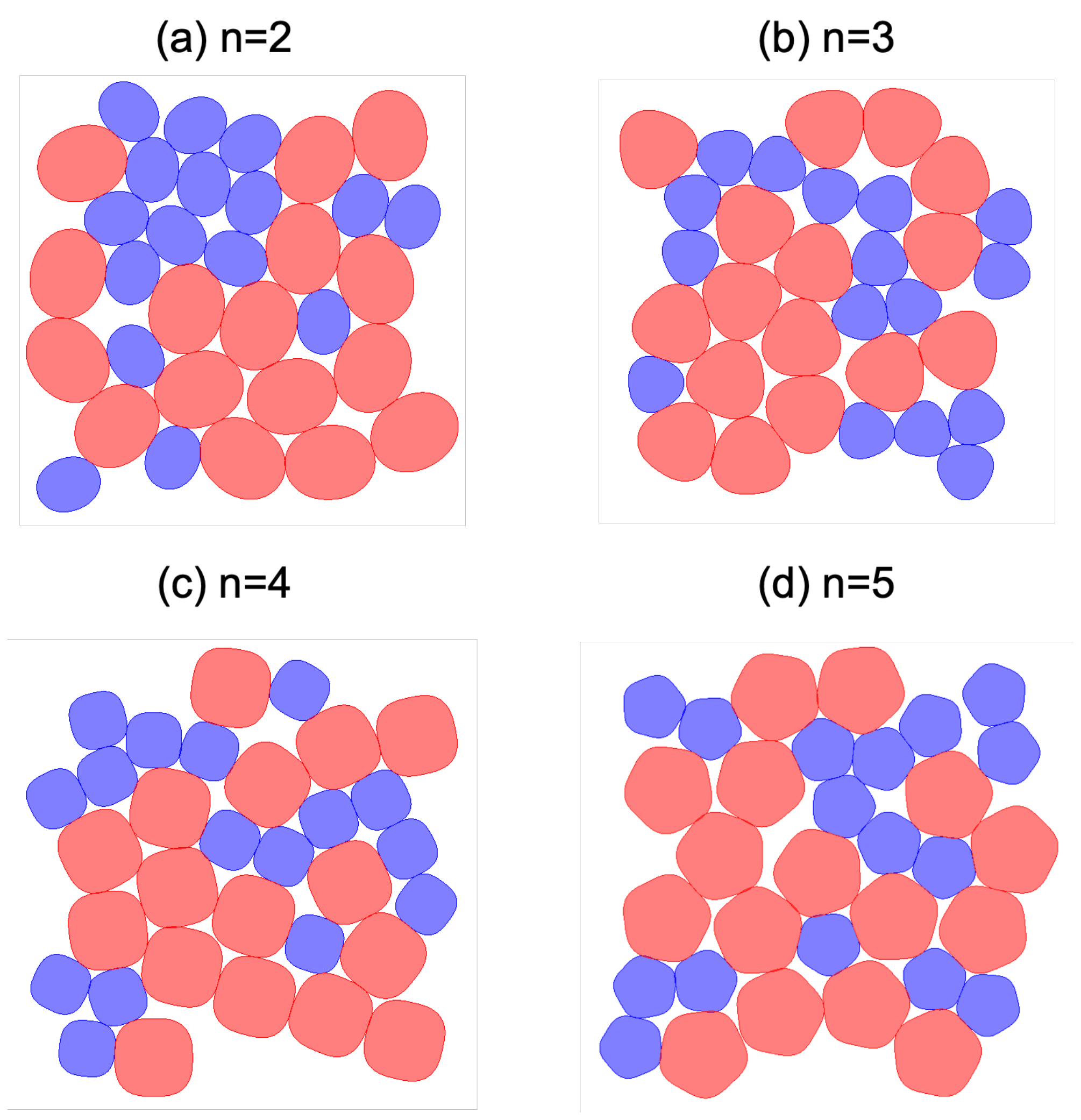} \caption{ Configurations at
 $\varphi_J$ for $N=32$, $\Delta=0.2$ and $n=2,\dots, 5$.}
 \label{140449_23Nov20}
\end{center}
\end{figure}

We perform numerical simulations for $N$
particles confined in a $L\times L$ box. We impose the periodic
boundary conditions for the both $x$ and $y$ directions. To avoid the
crystallization, we consider an equimolar binary mixture: $R_i^0=0.5$ for
$i=1,\dots, N/2$ and $R_i^0=0.7$ for $i=N/2+1,\dots, N$.  $\varphi_J$ is
the point at which $V_N$ begins to have a finite value. In practical, we
define $\varphi_J$ as a packing fraction satisfying~\cite{ohern2003}
\begin{align}
10^{-16} < V_N/N < 2\times 10^{-16}.
\end{align}
Here we explain how to generate configurations at $\varphi_J$. We first
generate a random initial configuration at small packing fraction $\varphi=0.1$.
Next, we slightly increase the density $\varphi \to \varphi +
\varepsilon$ with $\varepsilon = 10^{-3}$, and then minimize $V_N$ by
using the FIRE algorithm, which combines the standard molecular dynamics
of the Newton equation 
\begin{align}
 m\diff{^2\br_i}{t^2} = -\pdiff{V}{\br_i},\new
 I\diff{^2\theta_i}{t^2} = -\pdiff{V}{\theta_i},
\end{align}
with adaptive damping of the velocity~\cite{bitzek2006structural}. We
find that the FIRE converges in a reasonable time if we set $m=1$ and
$I=\Delta$ so that $\ddot{\br}_i=O(\Delta^0)$ and $\ddot{\theta}_i =
O(\Delta^0)$. We stop the FIRE algorithm when
$\left(\pdiff{V_N}{\br_i}\right)^2 +
\left(\pdiff{V_N}{\theta}_i\right)^2 < 10^{-25}$, or $V_N/N <
10^{-16}$~\cite{ohern2003}. We repeat the above compression
$\delta\varphi\to\varphi + \varepsilon$ and minimization protocols as
long as $V_N/N \leq 10^{-16}$ after the minimization. On the contrary,
if $V_N/N > 10^{-16}$ after the minimization, we then decompress the
system by changing the sign and amplitude of $\varepsilon$ as
$\varepsilon \to -\varepsilon/2$.  We repeat the above
compression/decompression protocols by changing $\varepsilon\to
-\varepsilon/2$ every time the energy crosses the threshold value
$V_N/N=10^{-16}$. We terminates the simulation when $10^{-16} < V_N/N <
2\times 10^{-16}$. In Fig.~\ref{140449_23Nov20}, we show configurations
at $\varphi_J$ generated by the above algorithm.
When calculate $z_J$ and $g(h)$, we remove the rattles, for which the
contact number is less than $d+1=3$~\cite{ohern2003}. To improve the
statistics, we take the average for $50$ independent samples.

\section{Comparison of our model and ellipses}
\label{132327_25Nov20}

\begin{figure}[t]
\begin{center}
 \includegraphics[width=9cm]{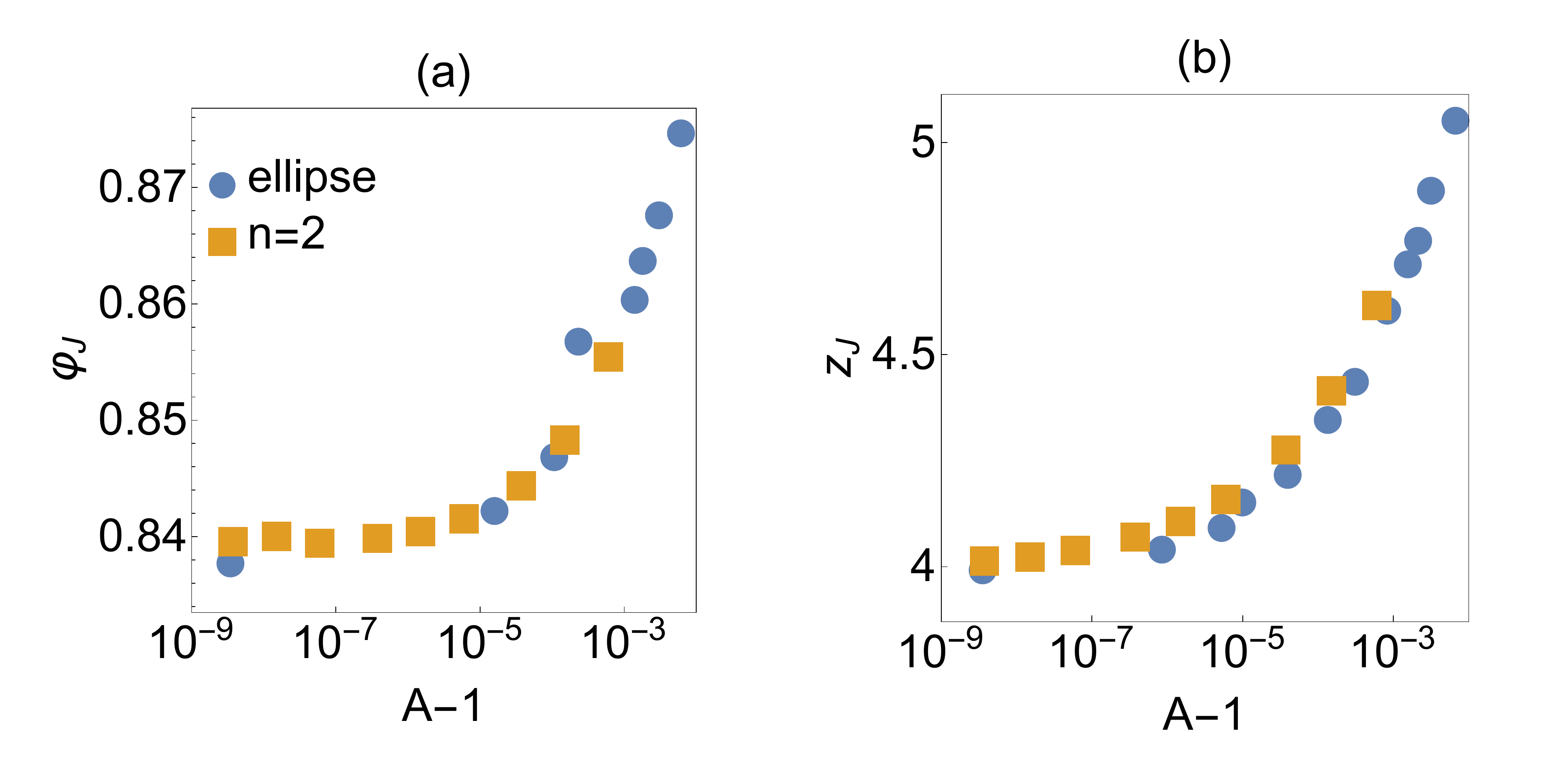} \caption{ Comparison of
 ellipses and our model for $n=2$. Data for ellipses are taken from
 Ref.~\cite{vander2018}.}  \label{055558_15Nov20}
\end{center}
\end{figure}

In this section, we compare our results with a previous numerical
simulation of ellipses~\cite{vander2018}, where $\varphi_J$ and $z_J$
were calculated as functions of $A$. The shape of an ellipse is defined
by the following equation:
\begin{align}
1  = \frac{x^2}{a^2} + \frac{y^2}{b^2}.
\end{align}
When the aspect ratio is close to one, say $a = R^0$ and $b =
R^0(1-\varepsilon)$ with $\varepsilon\ll 1$, we get
\begin{align}
R(\theta) &\equiv \sqrt{x^2 + y^2} \approx R^0\sqrt{1+2\varepsilon y^2 }\new
 &\approx R^0\left[1+\varepsilon + \varepsilon\sin\left(2\theta-\frac{\pi}{2}\right)
 \right].\label{055008_15Nov20}
\end{align}
As the constant term and phase shift do not change the shapes of
particles, Eq.~(\ref{055008_15Nov20}) suggests that ellipses can be
identified with our model of $n=2$ in the lowest order. In
Fig.~\ref{055558_15Nov20}, we compare our result for $n=2$ and $N=512$
with that of ellipses for $N=480$~\cite{vander2018} (the finite $N$
effect is not visible in the semi-log plot.). We find a reasonable
agreement for $A\sim 1$ or equivalently $\Delta\ll 1$, as expected.

\section{Packing fraction and fraction of rattles}
\label{133030_25Nov20}

\begin{figure}[t]
\begin{center}
 \includegraphics[width=8cm]{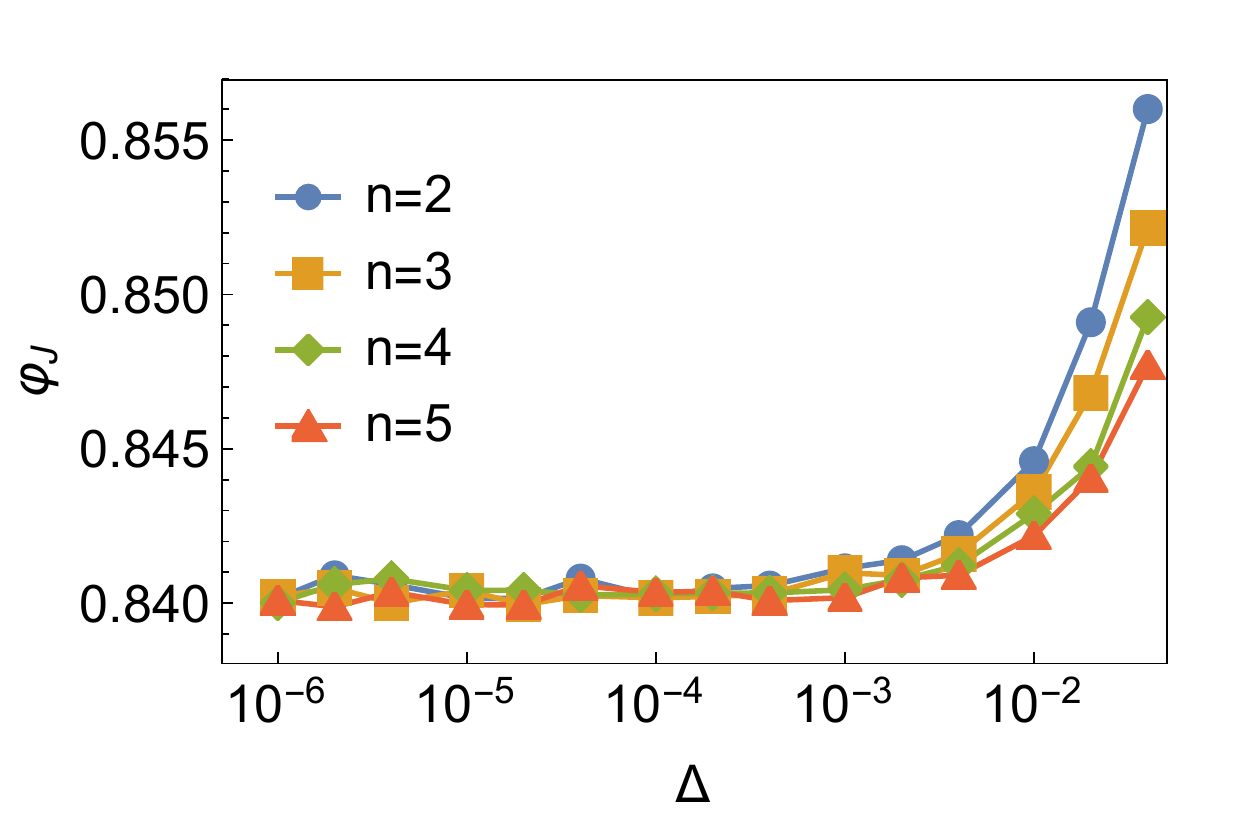} \caption{$\Delta$ dependence of
 $\varphi_J$ for $N=1024$.}  \label{122904_17Nov20}
\end{center}
\end{figure}
In Fig.~\ref{122904_17Nov20}, we show the $\Delta$ dependence of
$\varphi_J$ for $n=2,\dots, 5$. On increasing of $\Delta$, $\varphi_J$
increases for all $n$. This is consistent with the previous numerical
results of non-spherical particles, such as
ellipsoids~\cite{donev2004improving},
dimers~\cite{schreck2010comparison},
spherocylinders~\cite{williams2003random}, and
circulo-polygons~\cite{vander2018}. The more gradual increase of
$\varphi_J$ is observed for the larger $n$. It is interesting future
work to see what will happen for much larger value of $n$. For $n\to
\infty$, we expect that $\varphi_J$ \textit{decreases} on increasing of
$\Delta$, because the surface of particles becomes very rough and the
system may be mapped into frictional particles of the effective friction
coefficient $\mu \sim \abs{\partial_\theta V_N/\partial_{\br}V_N}\sim
\Delta$~\cite{tunable2020}.

\begin{figure}[t]
\begin{center}
 \includegraphics[width=8cm]{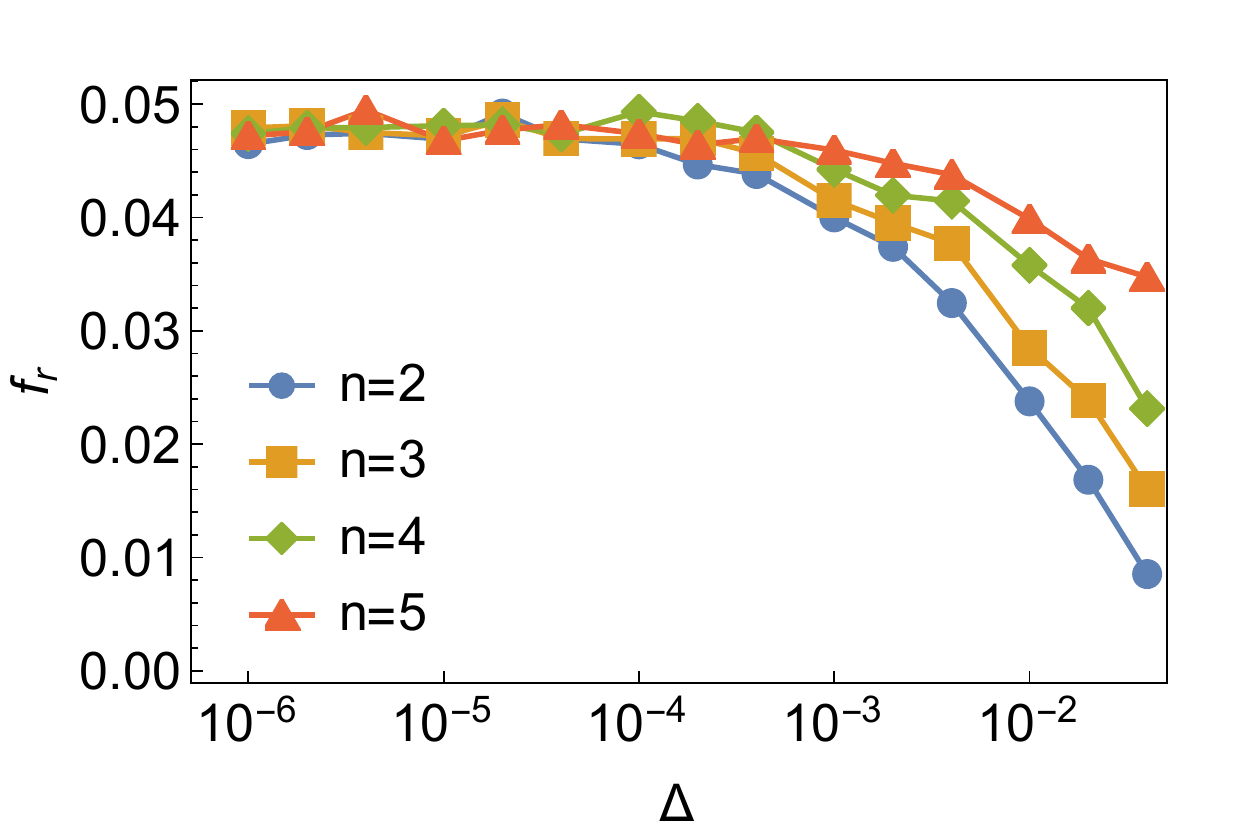} \caption{$\Delta$ dependence of
 the fraction of rattles $f_r$ for $N=1024$.}  \label{123546_17Nov20}
\end{center}
\end{figure}
In Fig.~\ref{123546_17Nov20}, we show the $\Delta$ dependence of the
fraction of rattles:
\begin{align}
 f_r = \frac{N_r}{N},
\end{align}
where $N_r$ denotes the number of rattles. We find that $f_r$ decreases
on increasing of $\Delta$ for all $n$. The similar results have been
reported for ellipses and ellipsoids~\cite{donev2007underconstrained}.

\section{Contact number at jamming}
\label{133113_25Nov20} In this section, we present our numerical results
for the contact number at the jamming transition point $z_J\equiv
z(\varphi_J)$.

\subsection{$N$ dependence for $n=2$}

\begin{figure}[t]
\begin{center}
 \includegraphics[width=9cm]{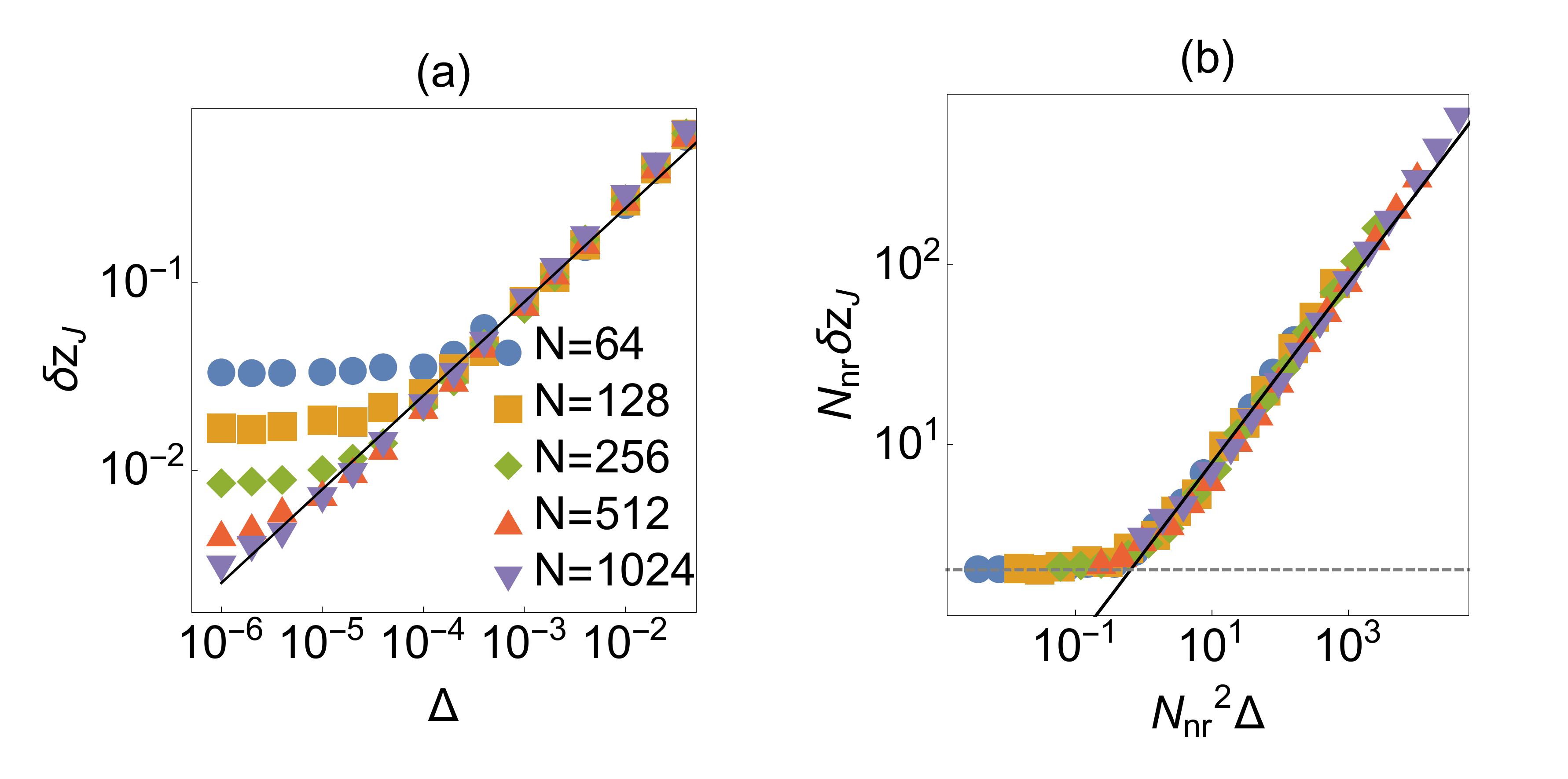} \caption{$\delta z_J$ for $n=2$
 and several $N$. (a) Markers denote the numerical results. (b) Scaling
 plot for the same data.  Black solid line and gray dashed line denote
 $\delta z_J\propto \Delta^{1/2}$ and $\delta z_J=2/N_{\rm nr}$,
 respectively } \label{142351_23Nov20}
\end{center}
\end{figure}

We first perform the finite size scaling analysis for $n=2$. In
Fig.~\ref{142351_23Nov20}, we show $\delta z_J=z_J-z_{\rm iso}$ of our
model for $n=2$ and $N=64,\dots, 1024$. We find a power-law region
$\delta z_J\sim \Delta^{1/2}$ for intermediate values of $\Delta$. The
power-law region becomes wider on increasing $N$.

Inspired by the finite $N$ scaling analysis for frictionless spherical
particles~\cite{goodrich2012}, we assume the following scaling
form:
\begin{align}
 \delta z_J \sim N_{\rm nr}^{-1}\tilde{z}(N_{\rm nr}^2\Delta),\label{172424_23Nov20}
\end{align}
where $N_{\rm nr}=N-N_r$ denotes the number of non-rattler particles,
and
\begin{align}
 \tilde{z}(x) 
\sim
\begin{cases}
 x^0 & x\ll 1\\
 x^{1/2} & x\gg 1\\
\end{cases}.\label{172417_23Nov20}
\end{align}
In Fig.~\ref{142351_23Nov20}~(b), we test the above scaling. A good
scaling collapse confirms Eq.~(\ref{172424_23Nov20}). Also, we find that
$\delta z_J \to 2/N_{\rm nr}$ for $N_{\rm nr}^2\Delta\to 0$, see the
dashed horizontal line in Fig.~\ref{142351_23Nov20}~(b). This means that
the system has just one extra contact than the number of degrees of
freedom, which is also consistent with the previous finite size analysis
of frictionless disks~\cite{goodrich2012}.


\subsection{$n$ dependence for $N=1024$}

\begin{figure}[t]
\begin{center}
 \includegraphics[width=8cm]{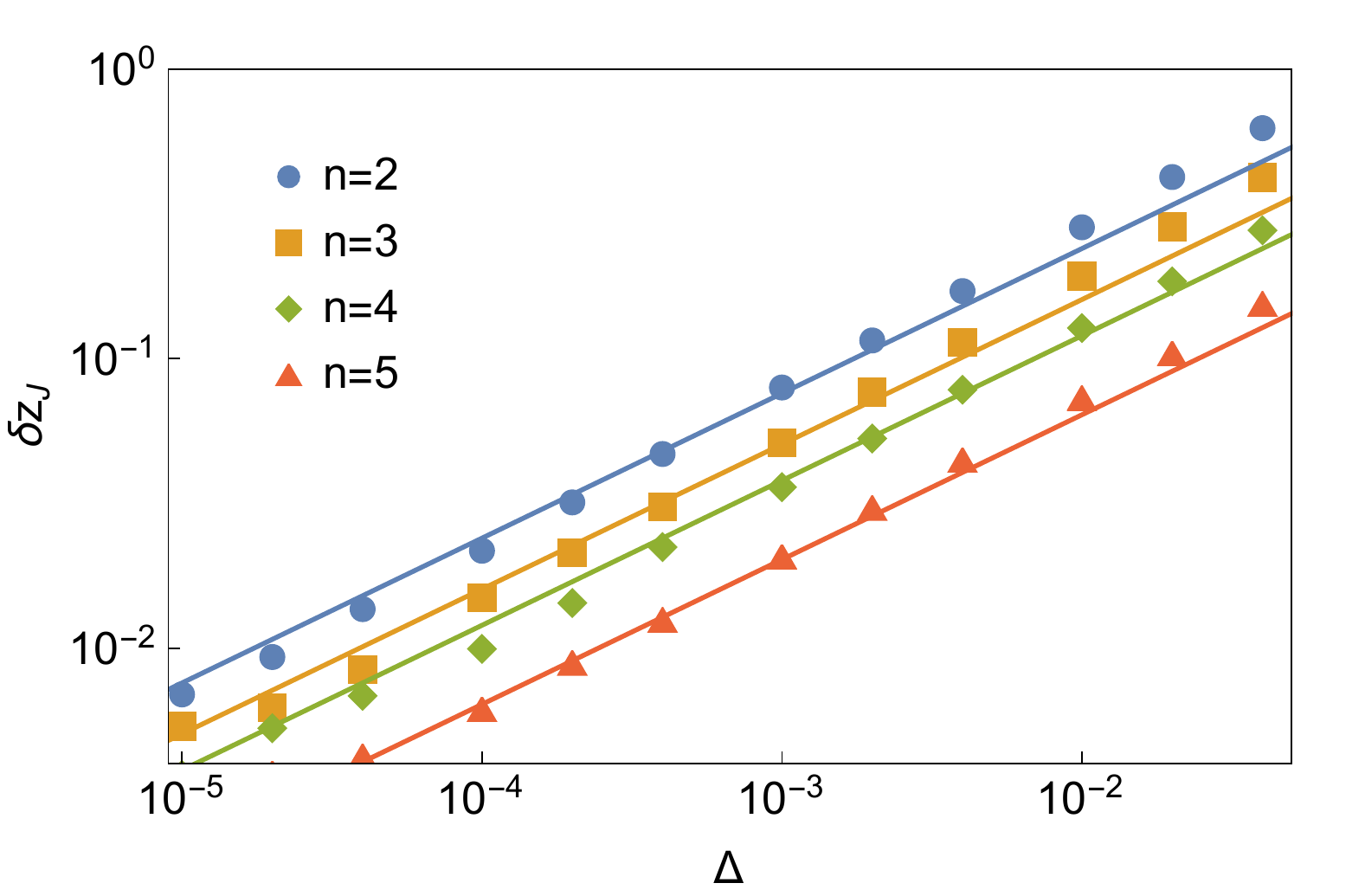} \caption{$\delta z_J$ for
 $N=1024$ and $n=2,\dots, 5$.  Markers denote the numerical results, and
 solid lines denote $\delta z_J\propto \Delta^{1/2}$.}
				       \label{172611_23Nov20}
\end{center}
\end{figure}
Now we focus on the data of the largest size $N=1024$. We only show the
results for $\delta z_J\gg 1/N$ so that the finite $N$ effects do not
appear.  In Fig.~\ref{172611_23Nov20}, we plot our numerical results of
$\delta z_J$ for $n=2,\dots, 5$. We find that $\delta z_J\propto
\Delta^{1/2}$ for all $n$, which confirms the mean-field prediction
Eq.~(\ref{152428_18Nov20}).

\section{Gap distribution}
\label{133141_25Nov20}

\begin{figure*}[t]
\begin{center}
 \includegraphics[width=18cm]{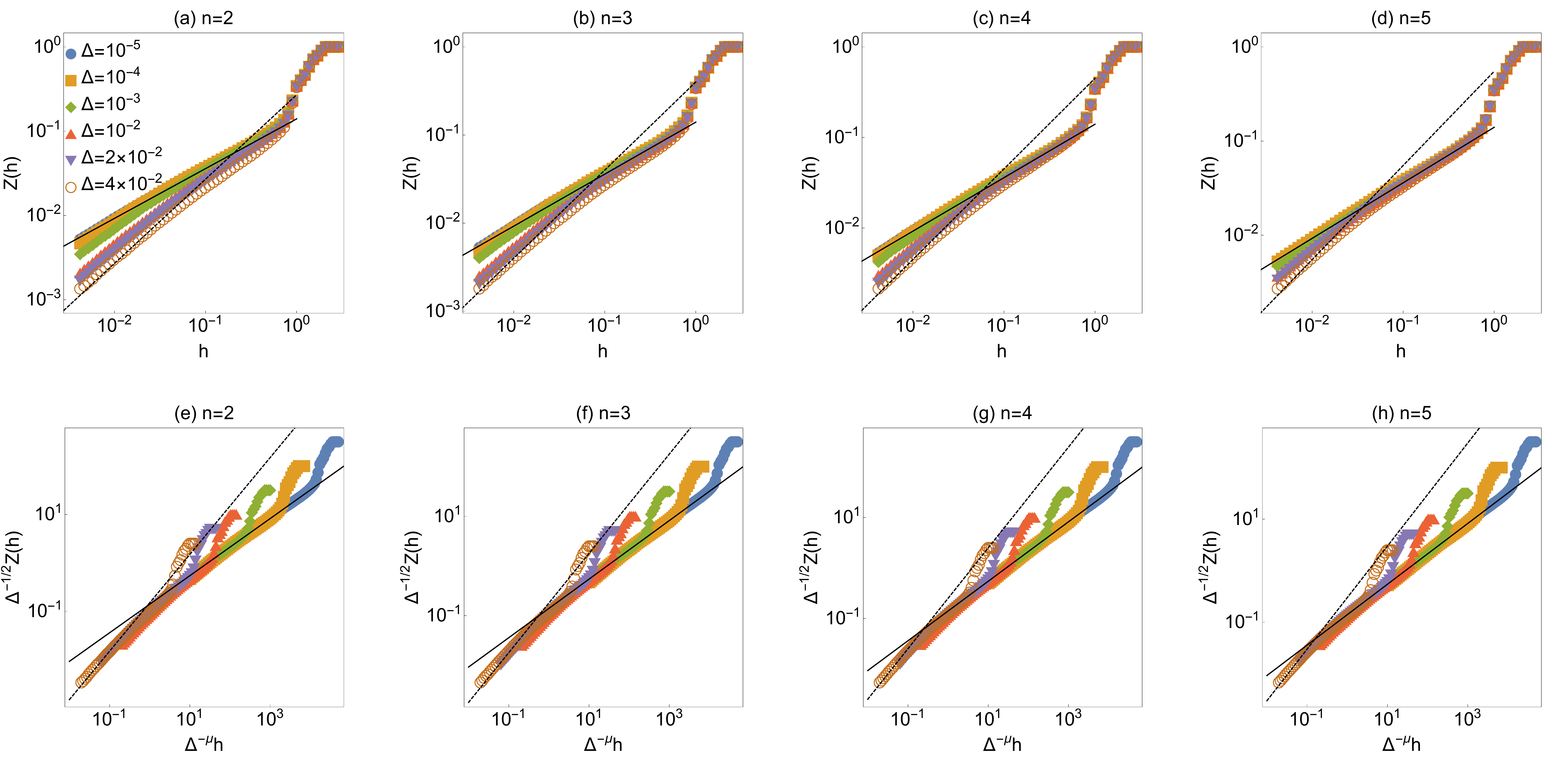} \caption{ (a--d) $Z(h)$ for
 $N=1024$, and $n=2,\dots, 5$. Markers denote the numerical results,
 solid line denotes $Z(h)\sim h^{1-\gamma}$, and dashed line denotes
 $Z(h)\sim h$.  (e--h) Scaling plot for the same data.}
					\label{150306_24Nov20}
\end{center}
\end{figure*}
In this section, we discuss the gap distribution $g(h)$ at
$\varphi_J$. To improve the statistics, instead of $g(h)$ itself, we
observe the cumulative distribution function:
\begin{align}
 Z(h) = \frac{\int_{0}^h g(h) dh}{\int_0^{h_{\rm cut}} g(h) dh}.
\end{align}
By definition $Z(0)=0$ and $Z(h_{\rm cut})=1$. We set $h_{\rm cut}=2$,
which is large enough to observe the scaling behavior.  In
Fig.~\ref{150306_24Nov20} (a--d), we show our numerical results of
$Z(h)$ for $n=2,\dots, 5$. We find that for small $\Delta$ and $h$,
$Z(h)$ exhibits the power-law $Z(h)\sim h^{1-\gamma}$, suggesting
$g(h)\sim h^{-\gamma}$.  On the contrary, for large $\Delta$, $Z(h)$
exhibits the liner behavior $Z(h)\sim h$ for $h\ll 1$, suggesting
$g(h)\sim h^{0}$.  These results are consistent with the mean-field
prediction Eq.~(\ref{152432_18Nov20}).

By using Eq.~(\ref{152432_18Nov20}), we can deduce the scaling form
of $Z(h)$ as~\cite{ikeda2020infinitesimal}
\begin{align}
 Z(h) \sim \Delta^{1/2}\tilde{Z}(\Delta^{-\mu}h),
\end{align}
where $\tilde{Z}(x)$ satisfies
\begin{align}
 \tilde{Z}(x) \sim
\begin{cases}
 x  & x\ll 1\\
 x^{1-\gamma} & x\gg 1
\end{cases}.
\end{align}
In Fig.~\ref{150306_24Nov20}~(e--h), we test the above equation.
We find a reasonable data collapse.

\section{Vibrational density of states}
\label{133202_25Nov20}

Finally, we investigate the vibrational density of states $D(\omega)$ at
$\varphi_J$. We define the Hessian of the interaction potential as
\begin{align}
& \HH_{X_i Y_j} = \pdiff{^2 V_N}{X_i \partial Y_j}
 = K_{X_i Y_j} + T_{X_i Y_j },\new
& K_{X_i Y_j} = \sum_{i<j}v''(h_{ij})\pdiff{h_{ij}}{X_i}\pdiff{h_{ij}}{Y_j},\new
& T_{X_i Y_j} = \sum_{i<j}v'(h_{ij})\pdiff{^2 h_{ij}}{X_i \partial Y_j},
\end{align}
where $X_i\in \{\br_i,\theta_i\}$ and $Y_j\in \{\br_j,\theta_j\}$. At
the jamming transition point, $v'(h_{ij})=0$, and thus
\begin{align}
&\HH_{X_i Y_j}\to K_{X_i Y_j}\new
& = \delta_{ij}\sum_{k\neq i}\Theta(-h_{ik})\pdiff{h_{ik}}{X_i}\pdiff{h_{ik}}{Y_i}
 + (1-\delta_{ij})\Theta(-h_{ij})\pdiff{h_{ij}}{X_i}\pdiff{h_{ij}}{Y_j}.
\end{align}
Using the eigenvalues of $H_{X_iY_j}$, $\{\lambda_n\}_{n=1,\dots, 3N}$,
$D(\omega)$ is calculated as
\begin{align}
 D(\omega) = \frac{1}{3N}\sum_{n=1}^{3N}\delta(\omega-\sqrt{\lambda_n}).
\end{align}
As mentioned below Eq.~(\ref{122941_23Nov20}), $D(\omega)$ has zero
modes at $\varphi_J$, in addition to the trivial zero modes related to
the rattler particles~\cite{mailman2009jamming,vander2018}. In
practice, however, the zero modes have finite frequencies depending on
the accuracy of the numerical simulation.  Hereafter, we focus on the
range $\omega>10^{-5}$, which is large enough to remove the zero modes.

\begin{figure}[t]
\begin{center}
 \includegraphics[width=9cm]{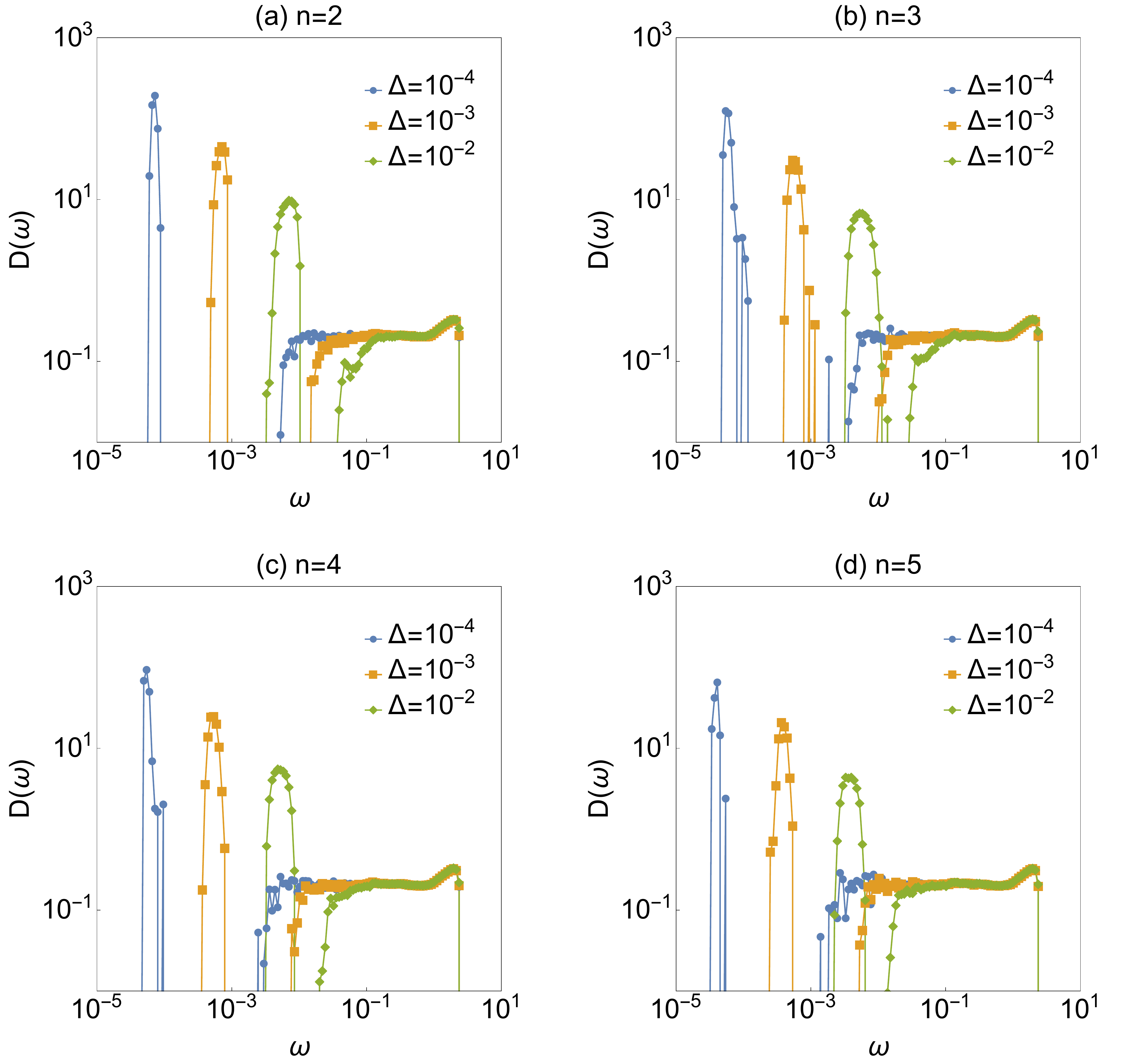} \caption{$D(\omega)$ for $N=1024$
 and $n=2,\dots, 5$. Here we do not show the zero modes.}
 \label{152933_24Nov20}
\end{center}
\end{figure}
In Fig.~\ref{152933_24Nov20}, we show our numerical results for
$D(\omega)$. We find that the behavior of the high $\omega$ region
($\omega > 0.1$) does not much depend on $\Delta$. On decreasing
$\Delta$, $D(\omega)$ develops a plateau down to the characteristic
frequency $\omega_*$. $D(\omega)$ has the separated band at $\omega_1\ll
\omega_*$. These results are consistent with the mean-field prediction
shown in Fig.~\ref{120311_23Nov20}~(b). Note that the lowest band in
Fig.~\ref{120311_23Nov20}~(b) does not appear, since $\omega_0=0$ at
$\varphi_J$, and we do not show the zero modes.
\begin{figure}[t]
\begin{center}
 \includegraphics[width=9cm]{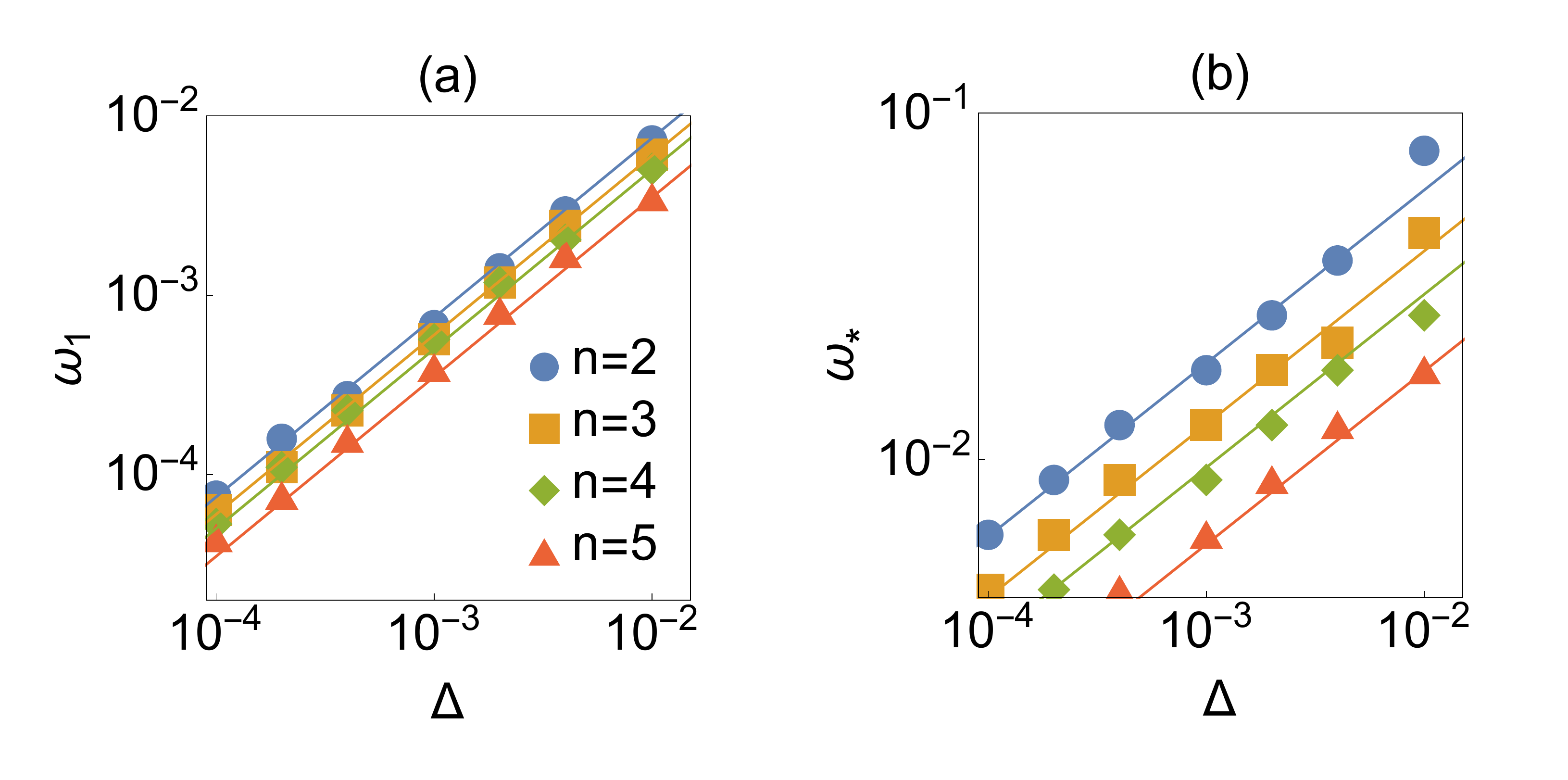} \caption{$\Delta$ dependence of
 characteristic frequencies. (a) Markers denote numerical results of
 $\omega_1$, while solid lines denote $\omega_1\sim\Delta$.  (b) (a)
 Markers denote numerical results of $\omega_*$, while solid lines
 denote $\omega_*\sim\Delta^{1/2}$.}  \label{161306_24Nov20}
\end{center}
\end{figure}
We want to calculate $\omega_1$ and $\omega_*$ from the numerical data
of $D(\omega)$. For this purpose, we define $\omega_1$ as the point that
maximizes $D(\omega)$, and $\omega_*$ as the point where
$D(\omega_*)=0.1$ in the range $\omega<0.1$. In
Fig.~\ref{161306_24Nov20}, we show $\Delta$ dependence of $\omega_1$ and
$\omega_*$. We find $\omega_1\sim \Delta$ and
$\omega_*\sim\Delta^{1/2}$, which are consistent with the mean-field
prediction, Eq.~(\ref{122941_23Nov20}). The similar results have been
previously reported for ellipses and
ellipsoids~\cite{schreck2012constraints,brito2018universality}.

\section{Summary and discussions}
\label{133229_25Nov20}

In this work, we performed a systematic numerical investigation for the
jamming of nearly spherical particles in two dimensions. Starting from
perfect disks, we systematically deformed the shapes of particles by the
$n$-th order term of the Fourier series $\propto\sin(n\theta)$ and
observed its effects on the physical quantities at the jamming
transition point. For an efficient numerical simulation, we derived an
analytic formula of the gap function by using the perturbation expansion
from the reference disks. By using the approximated gap function, we
numerically generated configurations at the jamming transition point,
and calculated the contact number, (cumulative) gap distribution, and
vibrational density of states for $n=2,\dots 5$. We found the
qualitatively the same scaling behaviors, which are fully consistent
with the mean-field predictions, for all $n$. This means that mean-field
prediction is applicable to general-shaped convex particles whose
particle shape can be represented by the Fourier series.

There are still several important points that deserve further
investigation.  Here we give a tentative list:
\begin{itemize}
 \item As mentioned before, our approximation does not hold for
       non-convex particles, such as dimers, where particles may have
       multiple contacts. It is important future work to extend the
       approximation for the gap function so as to take into account the
       effects of the multiple contacts.

 \item In this work, we investigate the physical quantities only at
       $\varphi_J$.  It is of course important to investigate the
       behavior for $\varphi>\varphi_J$. For instance, the mean-field
       theory predicts that the shear modulus $G$ behaves as
       \begin{align}
	G &\sim \Delta^{1/2}\tilde{G}\left(\Delta^{-1}\delta\varphi\right),
       \end{align}
       where $\tilde{G}(x)\sim x$ for $x\ll 1$ and $\tilde{G}(x)\sim
       x^{1/2}$ for $x\gg 1$~\cite{ikeda2020infinitesimal}. So far the
       above scaling is confirmed only for
       ellipsoids~\cite{ikeda2020infinitesimal}. It is important to test
       if the same scaling holds for other shapes of particles.
       
 \item The variational argument predicts that the correlation volume
       $v_{\rm corr}$ behaves as $v_{\rm corr}\sim \abs{\delta
       z^{-1}}$~\cite{yan2016variational}. For frictionless spherical
       particles, $\delta z=0$, thus $v_{\rm corr}$ diverges at
       $\varphi_J$. On the contrary, for non-spherical particles,
       $\delta z \neq 0$, therefore $v_{\rm corr}$ remains finite even
       at $\varphi_J$. Recently, it has been reported that $v_{\rm
       corr}$ can be extracted from the participation ratio of the
       lowest frequency mode of the vibrational density of
       states~\cite{shimada2018spatial}. It is interesting to repeat the
       same analysis for non-spherical particles.

 \item In this work, we focus on a system in two dimensions $d=2$. It is
       important future work to extend the current approximation and
       analysis to higher $d$.

 \item The mean-field theory of non-spherical particles predicts that
       the replica symmetry breaking (RSB) occurs near the jamming
       transition point~\cite{ikeda2019mean}, as in the case of
       spherical particles~\cite{charbonneau2014fractal}. It is 
       important future work to find out the signature of the RSB for
       non-spherical particles by numerical simulations and experiments.

 \item For frictionless spherical particles, the different critical
       exponents appear in the quasi-one-dimensional
       system~\cite{hikeda2020}. It is interesting future work to
       repeat the similar analysis in Ref.~~\cite{hikeda2020} for
       non-spherical particles.
\end{itemize}

\begin{acknowledgments}
This project has received funding from the JSPS KAKENHI Grant Number
JP20J00289.
\end{acknowledgments}

\appendix



\section{Isostaticity of particles consisting of spherical particles}
\label{075105_23Nov20}

To keep the generality, we consider $N$ particle system connected by $M$
bonds. For instance, $N/2$ dimers can be considered as $N$ spherical
particles with $M=N/2$ bonds. We consider the harmonic potential:
\begin{align}
V_N = \sum_{i<j}^{1,N}\frac{h_{ij}^2}{2}\Theta(-h_{ij})
 + k\sum_{a=1}^M \frac{t_{i_aj_a}^2}{2},\label{120957_27Nov20}
\end{align}
where
\begin{align}
& h_{ij} = \abs{\br_i-\br_j}-R_i-R_j,\new
& t_{i_aj_a} = \abs{\br_{i_a}-\br_{j_a}} - l_{i_a j_a}.
\end{align}
$\br_i=\{x_1^i, \dots, x_d^i\}$ denotes the position, and $R_i$ denotes
the diameter of the $i$-th particle, and $l_{i_aj_a}$ denotes the length
of the $a$-th bond connecting particles $i_a$ and $j_a$.

Here we show that the system is isostatic at $\varphi_J$ by using the
same argument for frictionless spherical
particles~\cite{wyart2005effects,wyart2005rigidity}.

The number of degrees of freedom of the system is
\begin{align}
 N_f = Nd.
\end{align}
At the jamming transition point, we have
\begin{align}
& \abs{\br_{i_\mu}-\br_{j_\mu}}= R_{i_\mu}+R_{i_\mu}, &\mu = 1,\dots, N_c,\new
& \abs{\br_{i_a}-\br_{j_a}}= l_{i_aj_a}, &a = 1,\dots, M.
 \label{081248_18Oct20}
\end{align}
where $N_c$ denotes the number of contacts, $i_\mu$ and $j_\mu$ denote
particles of the $\mu$-th contact.
One can find $\{\br_i\}_{i=1,\dots,
N}$ satisfying the above equation if
\begin{align}
Nd  \geq  N_{\rm const}, \label{100929_18Oct20}
\end{align}
where 
\begin{align}
 N_{\rm const} = N_c +M,
\end{align}
denotes the number of constraints at the jamming transition point.  On
the contrary, the force balance requires
\begin{align}
 \pdiff{V_N}{\br_i} = \sum_{j\neq i}\Theta(-h_{ij})h_{ij}\bn_{ij}
+ k\sum_{a=1}^M
 t_{ij_a}\bn_{ij_a} = 0,\label{095815_18Oct20}
\end{align}
where $\bn_{ij}$ denotes the normal vector connecting particles $i$ and
$j$. This can be regarded as $Nd$ linear equations for $\{h_{i_\mu
j_\mu}\}_{\mu=1,\dots, N_c}$ and $\{t_{i_aj_a}\}_{a=1,\dots, M}$. One
can find a solution if
\begin{align}
N_c+M \geq Nd.\label{100933_18Oct20}
\end{align}
From Eqs.~(\ref{100929_18Oct20}) and (\ref{100933_18Oct20}), we get
\begin{align}
 N_{\rm const} = N_f \leftrightarrow N_c + M = Nd,
\end{align}
meaning that the system is isostaticity at the jamming transition point.
For $N/2$ dimers, the total number of contacts 
is written as $N_c = (N/2)z_J/2$, leading to 
\begin{align}
 z_J = 4d-2,
\end{align}
which is consistent with the numerical results in
$d=2$~\cite{schreck2010comparison,shiraishi2019} and
$d=3$~\cite{shiraishi2020mechanical}.

\section{Derivation of Eq.~(\ref{084534_15Nov20})}
\label{131917_23Nov20}

\begin{figure}[t]
\begin{center}
 \includegraphics[width=7cm]{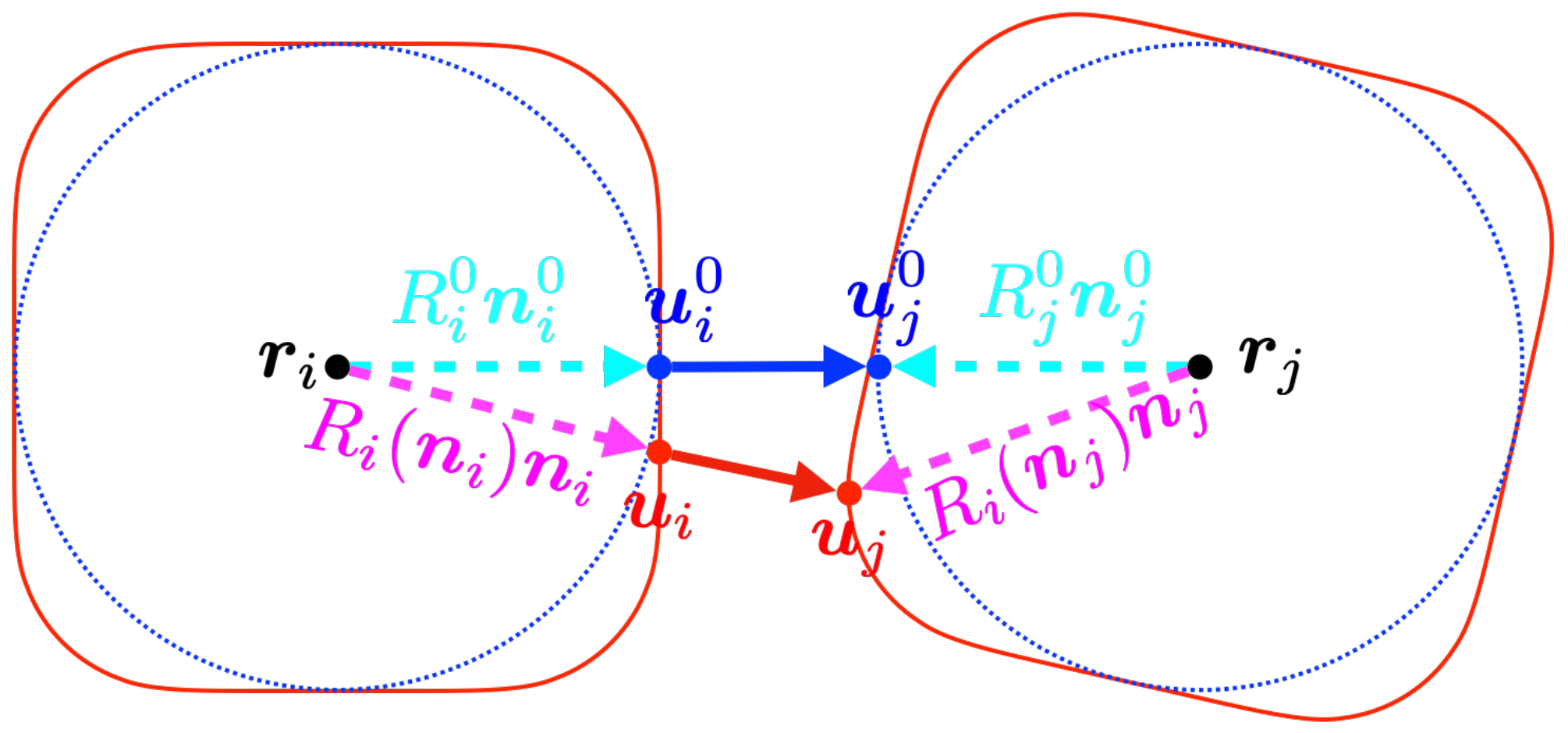} \caption{ Schematic picture of
 two non-spherical particles.  Red solid lines denote particles shape,
 and blue dashed lines denote reference disks.}  \label{055558_15Nov20}
\end{center}
\end{figure}

We write the gap function $h_{ij}$ as
\begin{align}
 h_{ij} &= \abs{\bu_i-\bu_j},
\end{align}
where $\bu_i$ and $\bu_j$ are points on the surfaces of particles $i$
and $j$ that minimize $h_{ij}$, see Fig.~\ref{055558_15Nov20}.  We
expand $\bu_i$ and $\bu_j$ from those of the reference disks as
\begin{align}
& \bu_i = \br_i + R_i(\bn_i)\bn_i
 = \bu_i^0 + \delta \bu_i,\new
& \bu_j = \br_j + R_j(\bn_j)\bn_j
 = \bu_j^0 + \delta \bu_j,
\end{align}
where
\begin{align}
&\bu_i^0 = \br_i + R_i^0\bn_i^0,\ \bu_j^0 = \br_j + R_j^0\bn_j^0,\new
& \delta \bu_i = R_i(\bn_i)\bn_i - R_i^0\bn_{i}^0,\delta \bu_j = R_j(\bn_j)\bn_j - R_j^0\bn_{j}^0,\new
&\bn_{i} = \frac{\br_i-\bu_i}{\abs{\br_i-\bu_i}},\bn_{j} = \frac{\br_j-\bu_j}{\abs{\br_j-\bu_j}},\new
&\bn_{i}^0 = \frac{\br_i-\br_j}{\abs{\br_i-\br_j}}, \bn_{j}^0 = -\bn_i^0,
\end{align}
see Fig.~\ref{055558_15Nov20}. $R_i(\bn_i)$ denotes the radius
of particle $i$ along the direction $\bn_i$.  In particular,
\begin{align}
 R_i(\bn_i^0) = R_i^0\left[1 + \Delta F(\theta_i-\theta_{ij})\right],
\end{align}
where $\theta_i$ denotes the direction of particle $i$, $\theta_{ij}$
denotes the relative angle between particles $i$ and $j$, see
Fig.~\ref{173935_24Nov20}.
For $\Delta\ll 1$,
we can expand $h_{ij}$ w.r.t $\Delta$
as 
\begin{align}
 h_{ij} &= \abs{\bu_i-\bu_j}
 = h_{ij}^0
 + \bn_i^0\cdot\left(\delta\bu_i - \delta\bu_j\right)
 + O(\delta\bu_i^2,\delta\bu_j^2)\new
 &= h_{ij}^0 + R_i^0-R_i(\bn_i)
 + R_j^0-R_j(\bn_j) + O(\Delta^2)\new
&=  h_{ij}^0 - \Delta \left[R_i^0 F(\theta_i-\theta_{ij})
 + R_j^0 F(\theta_j-\theta_{ji})\right] + O(\Delta^2),
\end{align}
where we used $\bn_i^0\cdot\bn_i = 1 + O(\Delta^2)$, and $R_i(\bn_i) =
R_i(\bn_i^0) + O(\Delta^2)$.  $h_{ij}^0$ denotes the gap function of the
reference disks:
\begin{align}
 h_{ij}^0 = \abs{\bu_i^0-\bu_j^0}
 = \abs{\br_i-\br_j}-R_i^0-R_j^0.
\end{align}


\bibliography{apssamp}

\begin{thebibliography}{47}%
\makeatletter
\providecommand \@ifxundefined [1]{%
 \@ifx{#1\undefined}
}%
\providecommand \@ifnum [1]{%
 \ifnum #1\expandafter \@firstoftwo
 \else \expandafter \@secondoftwo
 \fi
}%
\providecommand \@ifx [1]{%
 \ifx #1\expandafter \@firstoftwo
 \else \expandafter \@secondoftwo
 \fi
}%
\providecommand \natexlab [1]{#1}%
\providecommand \enquote  [1]{``#1''}%
\providecommand \bibnamefont  [1]{#1}%
\providecommand \bibfnamefont [1]{#1}%
\providecommand \citenamefont [1]{#1}%
\providecommand \href@noop [0]{\@secondoftwo}%
\providecommand \href [0]{\begingroup \@sanitize@url \@href}%
\providecommand \@href[1]{\@@startlink{#1}\@@href}%
\providecommand \@@href[1]{\endgroup#1\@@endlink}%
\providecommand \@sanitize@url [0]{\catcode `\\12\catcode `\$12\catcode
  `\&12\catcode `\#12\catcode `\^12\catcode `\_12\catcode `\%12\relax}%
\providecommand \@@startlink[1]{}%
\providecommand \@@endlink[0]{}%
\providecommand \url  [0]{\begingroup\@sanitize@url \@url }%
\providecommand \@url [1]{\endgroup\@href {#1}{\urlprefix }}%
\providecommand \urlprefix  [0]{URL }%
\providecommand \Eprint [0]{\href }%
\providecommand \doibase [0]{http://dx.doi.org/}%
\providecommand \selectlanguage [0]{\@gobble}%
\providecommand \bibinfo  [0]{\@secondoftwo}%
\providecommand \bibfield  [0]{\@secondoftwo}%
\providecommand \translation [1]{[#1]}%
\providecommand \BibitemOpen [0]{}%
\providecommand \bibitemStop [0]{}%
\providecommand \bibitemNoStop [0]{.\EOS\space}%
\providecommand \EOS [0]{\spacefactor3000\relax}%
\providecommand \BibitemShut  [1]{\csname bibitem#1\endcsname}%
\let\auto@bib@innerbib\@empty
\bibitem [{\citenamefont {van Hecke}(2009)}]{van2009jamming}%
  \BibitemOpen
  \bibfield  {author} {\bibinfo {author} {\bibfnamefont {M.}~\bibnamefont {van
  Hecke}},\ }\href@noop {} {\bibfield  {journal} {\bibinfo  {journal} {Journal
  of Physics: Condensed Matter}\ }\textbf {\bibinfo {volume} {22}},\ \bibinfo
  {pages} {033101} (\bibinfo {year} {2009})}\BibitemShut {NoStop}%
\bibitem [{\citenamefont {Liu}\ and\ \citenamefont
  {Nagel}(2010)}]{liu2010jamming}%
  \BibitemOpen
  \bibfield  {author} {\bibinfo {author} {\bibfnamefont {A.~J.}\ \bibnamefont
  {Liu}}\ and\ \bibinfo {author} {\bibfnamefont {S.~R.}\ \bibnamefont
  {Nagel}},\ }\href@noop {} {\bibfield  {journal} {\bibinfo  {journal} {Annu.
  Rev. Condens. Matter Phys.}\ }\textbf {\bibinfo {volume} {1}},\ \bibinfo
  {pages} {347} (\bibinfo {year} {2010})}\BibitemShut {NoStop}%
\bibitem [{\citenamefont {O'Hern}\ \emph {et~al.}(2003)\citenamefont {O'Hern},
  \citenamefont {Silbert}, \citenamefont {Liu},\ and\ \citenamefont
  {Nagel}}]{ohern2003}%
  \BibitemOpen
  \bibfield  {author} {\bibinfo {author} {\bibfnamefont {C.~S.}\ \bibnamefont
  {O'Hern}}, \bibinfo {author} {\bibfnamefont {L.~E.}\ \bibnamefont {Silbert}},
  \bibinfo {author} {\bibfnamefont {A.~J.}\ \bibnamefont {Liu}}, \ and\
  \bibinfo {author} {\bibfnamefont {S.~R.}\ \bibnamefont {Nagel}},\ }\href
  {\doibase 10.1103/PhysRevE.68.011306} {\bibfield  {journal} {\bibinfo
  {journal} {Phys. Rev. E}\ }\textbf {\bibinfo {volume} {68}},\ \bibinfo
  {pages} {011306} (\bibinfo {year} {2003})}\BibitemShut {NoStop}%
\bibitem [{\citenamefont {Nishimori}\ and\ \citenamefont
  {Ortiz}(2010)}]{nishimori2010elements}%
  \BibitemOpen
  \bibfield  {author} {\bibinfo {author} {\bibfnamefont {H.}~\bibnamefont
  {Nishimori}}\ and\ \bibinfo {author} {\bibfnamefont {G.}~\bibnamefont
  {Ortiz}},\ }\href@noop {} {\emph {\bibinfo {title} {Elements of phase
  transitions and critical phenomena}}}\ (\bibinfo  {publisher} {OUP Oxford},\
  \bibinfo {year} {2010})\BibitemShut {NoStop}%
\bibitem [{\citenamefont {V{\aa}gberg}\ \emph {et~al.}(2011)\citenamefont
  {V{\aa}gberg}, \citenamefont {Valdez-Balderas}, \citenamefont {Moore},
  \citenamefont {Olsson},\ and\ \citenamefont {Teitel}}]{vaagberg2011finite}%
  \BibitemOpen
  \bibfield  {author} {\bibinfo {author} {\bibfnamefont {D.}~\bibnamefont
  {V{\aa}gberg}}, \bibinfo {author} {\bibfnamefont {D.}~\bibnamefont
  {Valdez-Balderas}}, \bibinfo {author} {\bibfnamefont {M.}~\bibnamefont
  {Moore}}, \bibinfo {author} {\bibfnamefont {P.}~\bibnamefont {Olsson}}, \
  and\ \bibinfo {author} {\bibfnamefont {S.}~\bibnamefont {Teitel}},\
  }\href@noop {} {\bibfield  {journal} {\bibinfo  {journal} {Physical Review
  E}\ }\textbf {\bibinfo {volume} {83}},\ \bibinfo {pages} {030303} (\bibinfo
  {year} {2011})}\BibitemShut {NoStop}%
\bibitem [{\citenamefont {Charbonneau}\ \emph {et~al.}(2014)\citenamefont
  {Charbonneau}, \citenamefont {Kurchan}, \citenamefont {Parisi}, \citenamefont
  {Urbani},\ and\ \citenamefont {Zamponi}}]{charbonneau2014fractal}%
  \BibitemOpen
  \bibfield  {author} {\bibinfo {author} {\bibfnamefont {P.}~\bibnamefont
  {Charbonneau}}, \bibinfo {author} {\bibfnamefont {J.}~\bibnamefont
  {Kurchan}}, \bibinfo {author} {\bibfnamefont {G.}~\bibnamefont {Parisi}},
  \bibinfo {author} {\bibfnamefont {P.}~\bibnamefont {Urbani}}, \ and\ \bibinfo
  {author} {\bibfnamefont {F.}~\bibnamefont {Zamponi}},\ }\href@noop {}
  {\bibfield  {journal} {\bibinfo  {journal} {Nature communications}\ }\textbf
  {\bibinfo {volume} {5}},\ \bibinfo {pages} {1} (\bibinfo {year}
  {2014})}\BibitemShut {NoStop}%
\bibitem [{\citenamefont {Ikeda}(2020)}]{hikeda2020}%
  \BibitemOpen
  \bibfield  {author} {\bibinfo {author} {\bibfnamefont {H.}~\bibnamefont
  {Ikeda}},\ }\href {\doibase 10.1103/PhysRevLett.125.038001} {\bibfield
  {journal} {\bibinfo  {journal} {Phys. Rev. Lett.}\ }\textbf {\bibinfo
  {volume} {125}},\ \bibinfo {pages} {038001} (\bibinfo {year}
  {2020})}\BibitemShut {NoStop}%
\bibitem [{\citenamefont {Parisi}\ \emph {et~al.}(2020)\citenamefont {Parisi},
  \citenamefont {Urbani},\ and\ \citenamefont {Zamponi}}]{parisi2020theory}%
  \BibitemOpen
  \bibfield  {author} {\bibinfo {author} {\bibfnamefont {G.}~\bibnamefont
  {Parisi}}, \bibinfo {author} {\bibfnamefont {P.}~\bibnamefont {Urbani}}, \
  and\ \bibinfo {author} {\bibfnamefont {F.}~\bibnamefont {Zamponi}},\
  }\href@noop {} {\emph {\bibinfo {title} {Theory of simple glasses: exact
  solutions in infinite dimensions}}}\ (\bibinfo  {publisher} {Cambridge
  University Press},\ \bibinfo {year} {2020})\BibitemShut {NoStop}%
\bibitem [{\citenamefont {M{\'e}zard}\ \emph {et~al.}(1987)\citenamefont
  {M{\'e}zard}, \citenamefont {Parisi},\ and\ \citenamefont
  {Virasoro}}]{mezard1987spin}%
  \BibitemOpen
  \bibfield  {author} {\bibinfo {author} {\bibfnamefont {M.}~\bibnamefont
  {M{\'e}zard}}, \bibinfo {author} {\bibfnamefont {G.}~\bibnamefont {Parisi}},
  \ and\ \bibinfo {author} {\bibfnamefont {M.}~\bibnamefont {Virasoro}},\
  }\href@noop {} {\emph {\bibinfo {title} {Spin glass theory and beyond: An
  Introduction to the Replica Method and Its Applications}}},\ Vol.~\bibinfo
  {volume} {9}\ (\bibinfo  {publisher} {World Scientific Publishing Company},\
  \bibinfo {year} {1987})\BibitemShut {NoStop}%
\bibitem [{\citenamefont {Nishimori}(2001)}]{nishimori2001statistical}%
  \BibitemOpen
  \bibfield  {author} {\bibinfo {author} {\bibfnamefont {H.}~\bibnamefont
  {Nishimori}},\ }\href@noop {} {\emph {\bibinfo {title} {Statistical physics
  of spin glasses and information processing: an introduction}}},\ \bibinfo
  {number} {111}\ (\bibinfo  {publisher} {Clarendon Press},\ \bibinfo {year}
  {2001})\BibitemShut {NoStop}%
\bibitem [{\citenamefont {Wyart}\ \emph {et~al.}(2005)\citenamefont {Wyart},
  \citenamefont {Silbert}, \citenamefont {Nagel},\ and\ \citenamefont
  {Witten}}]{wyart2005effects}%
  \BibitemOpen
  \bibfield  {author} {\bibinfo {author} {\bibfnamefont {M.}~\bibnamefont
  {Wyart}}, \bibinfo {author} {\bibfnamefont {L.~E.}\ \bibnamefont {Silbert}},
  \bibinfo {author} {\bibfnamefont {S.~R.}\ \bibnamefont {Nagel}}, \ and\
  \bibinfo {author} {\bibfnamefont {T.~A.}\ \bibnamefont {Witten}},\
  }\href@noop {} {\bibfield  {journal} {\bibinfo  {journal} {Physical Review
  E}\ }\textbf {\bibinfo {volume} {72}},\ \bibinfo {pages} {051306} (\bibinfo
  {year} {2005})}\BibitemShut {NoStop}%
\bibitem [{\citenamefont {Yan}\ \emph {et~al.}(2016)\citenamefont {Yan},
  \citenamefont {DeGiuli},\ and\ \citenamefont {Wyart}}]{yan2016variational}%
  \BibitemOpen
  \bibfield  {author} {\bibinfo {author} {\bibfnamefont {L.}~\bibnamefont
  {Yan}}, \bibinfo {author} {\bibfnamefont {E.}~\bibnamefont {DeGiuli}}, \ and\
  \bibinfo {author} {\bibfnamefont {M.}~\bibnamefont {Wyart}},\ }\href@noop {}
  {\bibfield  {journal} {\bibinfo  {journal} {EPL (Europhysics Letters)}\
  }\textbf {\bibinfo {volume} {114}},\ \bibinfo {pages} {26003} (\bibinfo
  {year} {2016})}\BibitemShut {NoStop}%
\bibitem [{\citenamefont {DeGiuli}\ \emph
  {et~al.}(2014{\natexlab{a}})\citenamefont {DeGiuli}, \citenamefont
  {Laversanne-Finot}, \citenamefont {D{\"u}ring}, \citenamefont {Lerner},\ and\
  \citenamefont {Wyart}}]{degiuli2014effects}%
  \BibitemOpen
  \bibfield  {author} {\bibinfo {author} {\bibfnamefont {E.}~\bibnamefont
  {DeGiuli}}, \bibinfo {author} {\bibfnamefont {A.}~\bibnamefont
  {Laversanne-Finot}}, \bibinfo {author} {\bibfnamefont {G.}~\bibnamefont
  {D{\"u}ring}}, \bibinfo {author} {\bibfnamefont {E.}~\bibnamefont {Lerner}},
  \ and\ \bibinfo {author} {\bibfnamefont {M.}~\bibnamefont {Wyart}},\
  }\href@noop {} {\bibfield  {journal} {\bibinfo  {journal} {Soft Matter}\
  }\textbf {\bibinfo {volume} {10}},\ \bibinfo {pages} {5628} (\bibinfo {year}
  {2014}{\natexlab{a}})}\BibitemShut {NoStop}%
\bibitem [{\citenamefont {DeGiuli}\ \emph
  {et~al.}(2014{\natexlab{b}})\citenamefont {DeGiuli}, \citenamefont {Lerner},
  \citenamefont {Brito},\ and\ \citenamefont {Wyart}}]{degiuli2014force}%
  \BibitemOpen
  \bibfield  {author} {\bibinfo {author} {\bibfnamefont {E.}~\bibnamefont
  {DeGiuli}}, \bibinfo {author} {\bibfnamefont {E.}~\bibnamefont {Lerner}},
  \bibinfo {author} {\bibfnamefont {C.}~\bibnamefont {Brito}}, \ and\ \bibinfo
  {author} {\bibfnamefont {M.}~\bibnamefont {Wyart}},\ }\href@noop {}
  {\bibfield  {journal} {\bibinfo  {journal} {Proceedings of the National
  Academy of Sciences}\ }\textbf {\bibinfo {volume} {111}},\ \bibinfo {pages}
  {17054} (\bibinfo {year} {2014}{\natexlab{b}})}\BibitemShut {NoStop}%
\bibitem [{\citenamefont {Beltukov}(2015)}]{beltukov2015random}%
  \BibitemOpen
  \bibfield  {author} {\bibinfo {author} {\bibfnamefont {Y.}~\bibnamefont
  {Beltukov}},\ }\href@noop {} {\bibfield  {journal} {\bibinfo  {journal} {JETP
  Letters}\ }\textbf {\bibinfo {volume} {101}},\ \bibinfo {pages} {345}
  (\bibinfo {year} {2015})}\BibitemShut {NoStop}%
\bibitem [{\citenamefont {Ikeda}\ and\ \citenamefont
  {Shimada}(2020)}]{ikeda2020note}%
  \BibitemOpen
  \bibfield  {author} {\bibinfo {author} {\bibfnamefont {H.}~\bibnamefont
  {Ikeda}}\ and\ \bibinfo {author} {\bibfnamefont {M.}~\bibnamefont
  {Shimada}},\ }\href@noop {} {\bibfield  {journal} {\bibinfo  {journal} {arXiv
  preprint arXiv:2009.12060}\ } (\bibinfo {year} {2020})}\BibitemShut {NoStop}%
\bibitem [{\citenamefont {Brito}\ \emph {et~al.}(2018)\citenamefont {Brito},
  \citenamefont {Ikeda}, \citenamefont {Urbani}, \citenamefont {Wyart},\ and\
  \citenamefont {Zamponi}}]{brito2018universality}%
  \BibitemOpen
  \bibfield  {author} {\bibinfo {author} {\bibfnamefont {C.}~\bibnamefont
  {Brito}}, \bibinfo {author} {\bibfnamefont {H.}~\bibnamefont {Ikeda}},
  \bibinfo {author} {\bibfnamefont {P.}~\bibnamefont {Urbani}}, \bibinfo
  {author} {\bibfnamefont {M.}~\bibnamefont {Wyart}}, \ and\ \bibinfo {author}
  {\bibfnamefont {F.}~\bibnamefont {Zamponi}},\ }\href@noop {} {\bibfield
  {journal} {\bibinfo  {journal} {Proceedings of the National Academy of
  Sciences}\ }\textbf {\bibinfo {volume} {115}},\ \bibinfo {pages} {11736}
  (\bibinfo {year} {2018})}\BibitemShut {NoStop}%
\bibitem [{\citenamefont {Ikeda}\ \emph {et~al.}(2019)\citenamefont {Ikeda},
  \citenamefont {Urbani},\ and\ \citenamefont {Zamponi}}]{ikeda2019mean}%
  \BibitemOpen
  \bibfield  {author} {\bibinfo {author} {\bibfnamefont {H.}~\bibnamefont
  {Ikeda}}, \bibinfo {author} {\bibfnamefont {P.}~\bibnamefont {Urbani}}, \
  and\ \bibinfo {author} {\bibfnamefont {F.}~\bibnamefont {Zamponi}},\
  }\href@noop {} {\bibfield  {journal} {\bibinfo  {journal} {Journal of Physics
  A: Mathematical and Theoretical}\ }\textbf {\bibinfo {volume} {52}},\
  \bibinfo {pages} {344001} (\bibinfo {year} {2019})}\BibitemShut {NoStop}%
\bibitem [{\citenamefont {Lu}\ \emph {et~al.}(2015)\citenamefont {Lu},
  \citenamefont {Third},\ and\ \citenamefont {M{\"u}ller}}]{lu2015discrete}%
  \BibitemOpen
  \bibfield  {author} {\bibinfo {author} {\bibfnamefont {G.}~\bibnamefont
  {Lu}}, \bibinfo {author} {\bibfnamefont {J.}~\bibnamefont {Third}}, \ and\
  \bibinfo {author} {\bibfnamefont {C.}~\bibnamefont {M{\"u}ller}},\
  }\href@noop {} {\bibfield  {journal} {\bibinfo  {journal} {Chemical
  Engineering Science}\ }\textbf {\bibinfo {volume} {127}},\ \bibinfo {pages}
  {425} (\bibinfo {year} {2015})}\BibitemShut {NoStop}%
\bibitem [{\citenamefont {Ikeda}\ \emph
  {et~al.}(2020{\natexlab{a}})\citenamefont {Ikeda}, \citenamefont {Brito},\
  and\ \citenamefont {Wyart}}]{ikeda2020infinitesimal}%
  \BibitemOpen
  \bibfield  {author} {\bibinfo {author} {\bibfnamefont {H.}~\bibnamefont
  {Ikeda}}, \bibinfo {author} {\bibfnamefont {C.}~\bibnamefont {Brito}}, \ and\
  \bibinfo {author} {\bibfnamefont {M.}~\bibnamefont {Wyart}},\ }\href@noop {}
  {\bibfield  {journal} {\bibinfo  {journal} {Journal of Statistical Mechanics:
  Theory and Experiment}\ }\textbf {\bibinfo {volume} {2020}},\ \bibinfo
  {pages} {033302} (\bibinfo {year} {2020}{\natexlab{a}})}\BibitemShut
  {NoStop}%
\bibitem [{\citenamefont {Ikeda}\ \emph
  {et~al.}(2020{\natexlab{b}})\citenamefont {Ikeda}, \citenamefont {Brito},
  \citenamefont {Wyart},\ and\ \citenamefont {Zamponi}}]{tunable2020}%
  \BibitemOpen
  \bibfield  {author} {\bibinfo {author} {\bibfnamefont {H.}~\bibnamefont
  {Ikeda}}, \bibinfo {author} {\bibfnamefont {C.}~\bibnamefont {Brito}},
  \bibinfo {author} {\bibfnamefont {M.}~\bibnamefont {Wyart}}, \ and\ \bibinfo
  {author} {\bibfnamefont {F.}~\bibnamefont {Zamponi}},\ }\href {\doibase
  10.1103/PhysRevLett.124.208001} {\bibfield  {journal} {\bibinfo  {journal}
  {Phys. Rev. Lett.}\ }\textbf {\bibinfo {volume} {124}},\ \bibinfo {pages}
  {208001} (\bibinfo {year} {2020}{\natexlab{b}})}\BibitemShut {NoStop}%
\bibitem [{\citenamefont {Goodrich}\ \emph {et~al.}(2014)\citenamefont
  {Goodrich}, \citenamefont {Dagois-Bohy}, \citenamefont {Tighe}, \citenamefont
  {van Hecke}, \citenamefont {Liu},\ and\ \citenamefont
  {Nagel}}]{correction2011}%
  \BibitemOpen
  \bibfield  {author} {\bibinfo {author} {\bibfnamefont {C.~P.}\ \bibnamefont
  {Goodrich}}, \bibinfo {author} {\bibfnamefont {S.}~\bibnamefont
  {Dagois-Bohy}}, \bibinfo {author} {\bibfnamefont {B.~P.}\ \bibnamefont
  {Tighe}}, \bibinfo {author} {\bibfnamefont {M.}~\bibnamefont {van Hecke}},
  \bibinfo {author} {\bibfnamefont {A.~J.}\ \bibnamefont {Liu}}, \ and\
  \bibinfo {author} {\bibfnamefont {S.~R.}\ \bibnamefont {Nagel}},\ }\href
  {\doibase 10.1103/PhysRevE.90.022138} {\bibfield  {journal} {\bibinfo
  {journal} {Phys. Rev. E}\ }\textbf {\bibinfo {volume} {90}},\ \bibinfo
  {pages} {022138} (\bibinfo {year} {2014})}\BibitemShut {NoStop}%
\bibitem [{\citenamefont {Goodrich}\ \emph {et~al.}(2012)\citenamefont
  {Goodrich}, \citenamefont {Liu},\ and\ \citenamefont {Nagel}}]{goodrich2012}%
  \BibitemOpen
  \bibfield  {author} {\bibinfo {author} {\bibfnamefont {C.~P.}\ \bibnamefont
  {Goodrich}}, \bibinfo {author} {\bibfnamefont {A.~J.}\ \bibnamefont {Liu}}, \
  and\ \bibinfo {author} {\bibfnamefont {S.~R.}\ \bibnamefont {Nagel}},\ }\href
  {\doibase 10.1103/PhysRevLett.109.095704} {\bibfield  {journal} {\bibinfo
  {journal} {Phys. Rev. Lett.}\ }\textbf {\bibinfo {volume} {109}},\ \bibinfo
  {pages} {095704} (\bibinfo {year} {2012})}\BibitemShut {NoStop}%
\bibitem [{\citenamefont {Bernal}\ and\ \citenamefont
  {Mason}(1960)}]{bernal1960packing}%
  \BibitemOpen
  \bibfield  {author} {\bibinfo {author} {\bibfnamefont {J.}~\bibnamefont
  {Bernal}}\ and\ \bibinfo {author} {\bibfnamefont {J.}~\bibnamefont {Mason}},\
  }\href@noop {} {\bibfield  {journal} {\bibinfo  {journal} {Nature}\ }\textbf
  {\bibinfo {volume} {188}},\ \bibinfo {pages} {910} (\bibinfo {year}
  {1960})}\BibitemShut {NoStop}%
\bibitem [{\citenamefont {Donev}\ \emph {et~al.}(2005)\citenamefont {Donev},
  \citenamefont {Torquato},\ and\ \citenamefont {Stillinger}}]{donev2005}%
  \BibitemOpen
  \bibfield  {author} {\bibinfo {author} {\bibfnamefont {A.}~\bibnamefont
  {Donev}}, \bibinfo {author} {\bibfnamefont {S.}~\bibnamefont {Torquato}}, \
  and\ \bibinfo {author} {\bibfnamefont {F.~H.}\ \bibnamefont {Stillinger}},\
  }\href {\doibase 10.1103/PhysRevE.71.011105} {\bibfield  {journal} {\bibinfo
  {journal} {Phys. Rev. E}\ }\textbf {\bibinfo {volume} {71}},\ \bibinfo
  {pages} {011105} (\bibinfo {year} {2005})}\BibitemShut {NoStop}%
\bibitem [{\citenamefont {Charbonneau}\ \emph {et~al.}(2012)\citenamefont
  {Charbonneau}, \citenamefont {Corwin}, \citenamefont {Parisi},\ and\
  \citenamefont {Zamponi}}]{charbonneau2012}%
  \BibitemOpen
  \bibfield  {author} {\bibinfo {author} {\bibfnamefont {P.}~\bibnamefont
  {Charbonneau}}, \bibinfo {author} {\bibfnamefont {E.~I.}\ \bibnamefont
  {Corwin}}, \bibinfo {author} {\bibfnamefont {G.}~\bibnamefont {Parisi}}, \
  and\ \bibinfo {author} {\bibfnamefont {F.}~\bibnamefont {Zamponi}},\ }\href
  {\doibase 10.1103/PhysRevLett.109.205501} {\bibfield  {journal} {\bibinfo
  {journal} {Phys. Rev. Lett.}\ }\textbf {\bibinfo {volume} {109}},\ \bibinfo
  {pages} {205501} (\bibinfo {year} {2012})}\BibitemShut {NoStop}%
\bibitem [{\citenamefont {Charbonneau}\ \emph {et~al.}(2020)\citenamefont
  {Charbonneau}, \citenamefont {Corwin}, \citenamefont {Dennis}, \citenamefont
  {Rojas}, \citenamefont {Ikeda}, \citenamefont {Parisi},\ and\ \citenamefont
  {Ricci-Tersenghi}}]{charbonneau2020finite}%
  \BibitemOpen
  \bibfield  {author} {\bibinfo {author} {\bibfnamefont {P.}~\bibnamefont
  {Charbonneau}}, \bibinfo {author} {\bibfnamefont {E.}~\bibnamefont {Corwin}},
  \bibinfo {author} {\bibfnamefont {C.}~\bibnamefont {Dennis}}, \bibinfo
  {author} {\bibfnamefont {R.~D.~H.}\ \bibnamefont {Rojas}}, \bibinfo {author}
  {\bibfnamefont {H.}~\bibnamefont {Ikeda}}, \bibinfo {author} {\bibfnamefont
  {G.}~\bibnamefont {Parisi}}, \ and\ \bibinfo {author} {\bibfnamefont
  {F.}~\bibnamefont {Ricci-Tersenghi}},\ }\href@noop {} {\bibfield  {journal}
  {\bibinfo  {journal} {arXiv preprint arXiv:2011.10899}\ } (\bibinfo {year}
  {2020})}\BibitemShut {NoStop}%
\bibitem [{\citenamefont {Franz}\ \emph {et~al.}(2017)\citenamefont {Franz},
  \citenamefont {Parisi}, \citenamefont {Sevelev}, \citenamefont {Urbani},\
  and\ \citenamefont {Zamponi}}]{franz2017universality}%
  \BibitemOpen
  \bibfield  {author} {\bibinfo {author} {\bibfnamefont {S.}~\bibnamefont
  {Franz}}, \bibinfo {author} {\bibfnamefont {G.}~\bibnamefont {Parisi}},
  \bibinfo {author} {\bibfnamefont {M.}~\bibnamefont {Sevelev}}, \bibinfo
  {author} {\bibfnamefont {P.}~\bibnamefont {Urbani}}, \ and\ \bibinfo {author}
  {\bibfnamefont {F.}~\bibnamefont {Zamponi}},\ }\href@noop {} {\  (\bibinfo
  {year} {2017})}\BibitemShut {NoStop}%
\bibitem [{\citenamefont {Schreck}\ \emph {et~al.}(2010)\citenamefont
  {Schreck}, \citenamefont {Xu},\ and\ \citenamefont
  {O'Hern}}]{schreck2010comparison}%
  \BibitemOpen
  \bibfield  {author} {\bibinfo {author} {\bibfnamefont {C.~F.}\ \bibnamefont
  {Schreck}}, \bibinfo {author} {\bibfnamefont {N.}~\bibnamefont {Xu}}, \ and\
  \bibinfo {author} {\bibfnamefont {C.~S.}\ \bibnamefont {O'Hern}},\
  }\href@noop {} {\bibfield  {journal} {\bibinfo  {journal} {Soft Matter}\
  }\textbf {\bibinfo {volume} {6}},\ \bibinfo {pages} {2960} (\bibinfo {year}
  {2010})}\BibitemShut {NoStop}%
\bibitem [{\citenamefont {Shiraishi}\ \emph {et~al.}(2019)\citenamefont
  {Shiraishi}, \citenamefont {Mizuno},\ and\ \citenamefont
  {Ikeda}}]{shiraishi2019}%
  \BibitemOpen
  \bibfield  {author} {\bibinfo {author} {\bibfnamefont {K.}~\bibnamefont
  {Shiraishi}}, \bibinfo {author} {\bibfnamefont {H.}~\bibnamefont {Mizuno}}, \
  and\ \bibinfo {author} {\bibfnamefont {A.}~\bibnamefont {Ikeda}},\ }\href
  {\doibase 10.1103/PhysRevE.100.012606} {\bibfield  {journal} {\bibinfo
  {journal} {Phys. Rev. E}\ }\textbf {\bibinfo {volume} {100}},\ \bibinfo
  {pages} {012606} (\bibinfo {year} {2019})}\BibitemShut {NoStop}%
\bibitem [{\citenamefont {Shiraishi}\ \emph {et~al.}(2020)\citenamefont
  {Shiraishi}, \citenamefont {Mizuno},\ and\ \citenamefont
  {Ikeda}}]{shiraishi2020mechanical}%
  \BibitemOpen
  \bibfield  {author} {\bibinfo {author} {\bibfnamefont {K.}~\bibnamefont
  {Shiraishi}}, \bibinfo {author} {\bibfnamefont {H.}~\bibnamefont {Mizuno}}, \
  and\ \bibinfo {author} {\bibfnamefont {A.}~\bibnamefont {Ikeda}},\
  }\href@noop {} {\bibfield  {journal} {\bibinfo  {journal} {arXiv preprint
  arXiv:2005.02598}\ } (\bibinfo {year} {2020})}\BibitemShut {NoStop}%
\bibitem [{\citenamefont {Donev}\ \emph {et~al.}(2004)\citenamefont {Donev},
  \citenamefont {Cisse}, \citenamefont {Sachs}, \citenamefont {Variano},
  \citenamefont {Stillinger}, \citenamefont {Connelly}, \citenamefont
  {Torquato},\ and\ \citenamefont {Chaikin}}]{donev2004improving}%
  \BibitemOpen
  \bibfield  {author} {\bibinfo {author} {\bibfnamefont {A.}~\bibnamefont
  {Donev}}, \bibinfo {author} {\bibfnamefont {I.}~\bibnamefont {Cisse}},
  \bibinfo {author} {\bibfnamefont {D.}~\bibnamefont {Sachs}}, \bibinfo
  {author} {\bibfnamefont {E.~A.}\ \bibnamefont {Variano}}, \bibinfo {author}
  {\bibfnamefont {F.~H.}\ \bibnamefont {Stillinger}}, \bibinfo {author}
  {\bibfnamefont {R.}~\bibnamefont {Connelly}}, \bibinfo {author}
  {\bibfnamefont {S.}~\bibnamefont {Torquato}}, \ and\ \bibinfo {author}
  {\bibfnamefont {P.~M.}\ \bibnamefont {Chaikin}},\ }\href@noop {} {\bibfield
  {journal} {\bibinfo  {journal} {Science}\ }\textbf {\bibinfo {volume}
  {303}},\ \bibinfo {pages} {990} (\bibinfo {year} {2004})}\BibitemShut
  {NoStop}%
\bibitem [{\citenamefont {Donev}\ \emph {et~al.}(2007)\citenamefont {Donev},
  \citenamefont {Connelly}, \citenamefont {Stillinger},\ and\ \citenamefont
  {Torquato}}]{donev2007underconstrained}%
  \BibitemOpen
  \bibfield  {author} {\bibinfo {author} {\bibfnamefont {A.}~\bibnamefont
  {Donev}}, \bibinfo {author} {\bibfnamefont {R.}~\bibnamefont {Connelly}},
  \bibinfo {author} {\bibfnamefont {F.~H.}\ \bibnamefont {Stillinger}}, \ and\
  \bibinfo {author} {\bibfnamefont {S.}~\bibnamefont {Torquato}},\ }\href@noop
  {} {\bibfield  {journal} {\bibinfo  {journal} {Physical Review E}\ }\textbf
  {\bibinfo {volume} {75}},\ \bibinfo {pages} {051304} (\bibinfo {year}
  {2007})}\BibitemShut {NoStop}%
\bibitem [{\citenamefont {Zeravcic}\ \emph {et~al.}(2009)\citenamefont
  {Zeravcic}, \citenamefont {Xu}, \citenamefont {Liu}, \citenamefont {Nagel},\
  and\ \citenamefont {van Saarloos}}]{zeravcic2009excitations}%
  \BibitemOpen
  \bibfield  {author} {\bibinfo {author} {\bibfnamefont {Z.}~\bibnamefont
  {Zeravcic}}, \bibinfo {author} {\bibfnamefont {N.}~\bibnamefont {Xu}},
  \bibinfo {author} {\bibfnamefont {A.}~\bibnamefont {Liu}}, \bibinfo {author}
  {\bibfnamefont {S.}~\bibnamefont {Nagel}}, \ and\ \bibinfo {author}
  {\bibfnamefont {W.}~\bibnamefont {van Saarloos}},\ }\href@noop {} {\bibfield
  {journal} {\bibinfo  {journal} {EPL (Europhysics Letters)}\ }\textbf
  {\bibinfo {volume} {87}},\ \bibinfo {pages} {26001} (\bibinfo {year}
  {2009})}\BibitemShut {NoStop}%
\bibitem [{\citenamefont {Mailman}\ \emph {et~al.}(2009)\citenamefont
  {Mailman}, \citenamefont {Schreck}, \citenamefont {O’Hern},\ and\
  \citenamefont {Chakraborty}}]{mailman2009jamming}%
  \BibitemOpen
  \bibfield  {author} {\bibinfo {author} {\bibfnamefont {M.}~\bibnamefont
  {Mailman}}, \bibinfo {author} {\bibfnamefont {C.~F.}\ \bibnamefont
  {Schreck}}, \bibinfo {author} {\bibfnamefont {C.~S.}\ \bibnamefont
  {O’Hern}}, \ and\ \bibinfo {author} {\bibfnamefont {B.}~\bibnamefont
  {Chakraborty}},\ }\href@noop {} {\bibfield  {journal} {\bibinfo  {journal}
  {Physical review letters}\ }\textbf {\bibinfo {volume} {102}},\ \bibinfo
  {pages} {255501} (\bibinfo {year} {2009})}\BibitemShut {NoStop}%
\bibitem [{\citenamefont {Schreck}\ \emph {et~al.}(2012)\citenamefont
  {Schreck}, \citenamefont {Mailman}, \citenamefont {Chakraborty},\ and\
  \citenamefont {O'Hern}}]{schreck2012constraints}%
  \BibitemOpen
  \bibfield  {author} {\bibinfo {author} {\bibfnamefont {C.~F.}\ \bibnamefont
  {Schreck}}, \bibinfo {author} {\bibfnamefont {M.}~\bibnamefont {Mailman}},
  \bibinfo {author} {\bibfnamefont {B.}~\bibnamefont {Chakraborty}}, \ and\
  \bibinfo {author} {\bibfnamefont {C.~S.}\ \bibnamefont {O'Hern}},\
  }\href@noop {} {\bibfield  {journal} {\bibinfo  {journal} {Physical Review
  E}\ }\textbf {\bibinfo {volume} {85}},\ \bibinfo {pages} {061305} (\bibinfo
  {year} {2012})}\BibitemShut {NoStop}%
\bibitem [{\citenamefont {Williams}\ and\ \citenamefont
  {Philipse}(2003)}]{williams2003random}%
  \BibitemOpen
  \bibfield  {author} {\bibinfo {author} {\bibfnamefont {S.}~\bibnamefont
  {Williams}}\ and\ \bibinfo {author} {\bibfnamefont {A.}~\bibnamefont
  {Philipse}},\ }\href@noop {} {\bibfield  {journal} {\bibinfo  {journal}
  {Physical Review E}\ }\textbf {\bibinfo {volume} {67}},\ \bibinfo {pages}
  {051301} (\bibinfo {year} {2003})}\BibitemShut {NoStop}%
\bibitem [{\citenamefont {Blouwolff}\ and\ \citenamefont
  {Fraden}(2006)}]{blouwolff2006coordination}%
  \BibitemOpen
  \bibfield  {author} {\bibinfo {author} {\bibfnamefont {J.}~\bibnamefont
  {Blouwolff}}\ and\ \bibinfo {author} {\bibfnamefont {S.}~\bibnamefont
  {Fraden}},\ }\href@noop {} {\bibfield  {journal} {\bibinfo  {journal} {EPL
  (Europhysics Letters)}\ }\textbf {\bibinfo {volume} {76}},\ \bibinfo {pages}
  {1095} (\bibinfo {year} {2006})}\BibitemShut {NoStop}%
\bibitem [{\citenamefont {Az\'ema}\ and\ \citenamefont
  {Radja\"{\i}}(2010)}]{stress2010}%
  \BibitemOpen
  \bibfield  {author} {\bibinfo {author} {\bibfnamefont {E.}~\bibnamefont
  {Az\'ema}}\ and\ \bibinfo {author} {\bibfnamefont {F.}~\bibnamefont
  {Radja\"{\i}}},\ }\href {\doibase 10.1103/PhysRevE.81.051304} {\bibfield
  {journal} {\bibinfo  {journal} {Phys. Rev. E}\ }\textbf {\bibinfo {volume}
  {81}},\ \bibinfo {pages} {051304} (\bibinfo {year} {2010})}\BibitemShut
  {NoStop}%
\bibitem [{\citenamefont {Marschall}\ and\ \citenamefont
  {Teitel}(2018)}]{marschall2018compression}%
  \BibitemOpen
  \bibfield  {author} {\bibinfo {author} {\bibfnamefont {T.}~\bibnamefont
  {Marschall}}\ and\ \bibinfo {author} {\bibfnamefont {S.}~\bibnamefont
  {Teitel}},\ }\href@noop {} {\bibfield  {journal} {\bibinfo  {journal}
  {Physical Review E}\ }\textbf {\bibinfo {volume} {97}},\ \bibinfo {pages}
  {012905} (\bibinfo {year} {2018})}\BibitemShut {NoStop}%
\bibitem [{\citenamefont {Jiao}\ \emph {et~al.}(2010)\citenamefont {Jiao},
  \citenamefont {Stillinger},\ and\ \citenamefont {Torquato}}]{jiao2010}%
  \BibitemOpen
  \bibfield  {author} {\bibinfo {author} {\bibfnamefont {Y.}~\bibnamefont
  {Jiao}}, \bibinfo {author} {\bibfnamefont {F.~H.}\ \bibnamefont
  {Stillinger}}, \ and\ \bibinfo {author} {\bibfnamefont {S.}~\bibnamefont
  {Torquato}},\ }\href {\doibase 10.1103/PhysRevE.81.041304} {\bibfield
  {journal} {\bibinfo  {journal} {Phys. Rev. E}\ }\textbf {\bibinfo {volume}
  {81}},\ \bibinfo {pages} {041304} (\bibinfo {year} {2010})}\BibitemShut
  {NoStop}%
\bibitem [{\citenamefont {Delaney}\ and\ \citenamefont
  {Cleary}(2010)}]{delaney2010packing}%
  \BibitemOpen
  \bibfield  {author} {\bibinfo {author} {\bibfnamefont {G.~W.}\ \bibnamefont
  {Delaney}}\ and\ \bibinfo {author} {\bibfnamefont {P.~W.}\ \bibnamefont
  {Cleary}},\ }\href@noop {} {\bibfield  {journal} {\bibinfo  {journal} {EPL
  (Europhysics Letters)}\ }\textbf {\bibinfo {volume} {89}},\ \bibinfo {pages}
  {34002} (\bibinfo {year} {2010})}\BibitemShut {NoStop}%
\bibitem [{\citenamefont {VanderWerf}\ \emph {et~al.}(2018)\citenamefont
  {VanderWerf}, \citenamefont {Jin}, \citenamefont {Shattuck},\ and\
  \citenamefont {O'Hern}}]{vander2018}%
  \BibitemOpen
  \bibfield  {author} {\bibinfo {author} {\bibfnamefont {K.}~\bibnamefont
  {VanderWerf}}, \bibinfo {author} {\bibfnamefont {W.}~\bibnamefont {Jin}},
  \bibinfo {author} {\bibfnamefont {M.~D.}\ \bibnamefont {Shattuck}}, \ and\
  \bibinfo {author} {\bibfnamefont {C.~S.}\ \bibnamefont {O'Hern}},\ }\href
  {\doibase 10.1103/PhysRevE.97.012909} {\bibfield  {journal} {\bibinfo
  {journal} {Phys. Rev. E}\ }\textbf {\bibinfo {volume} {97}},\ \bibinfo
  {pages} {012909} (\bibinfo {year} {2018})}\BibitemShut {NoStop}%
\bibitem [{\citenamefont {Tarama}\ \emph {et~al.}(2013)\citenamefont {Tarama},
  \citenamefont {Menzel}, \citenamefont {ten Hagen}, \citenamefont
  {Wittkowski}, \citenamefont {Ohta},\ and\ \citenamefont
  {L{\"o}wen}}]{tarama2013dynamics}%
  \BibitemOpen
  \bibfield  {author} {\bibinfo {author} {\bibfnamefont {M.}~\bibnamefont
  {Tarama}}, \bibinfo {author} {\bibfnamefont {A.~M.}\ \bibnamefont {Menzel}},
  \bibinfo {author} {\bibfnamefont {B.}~\bibnamefont {ten Hagen}}, \bibinfo
  {author} {\bibfnamefont {R.}~\bibnamefont {Wittkowski}}, \bibinfo {author}
  {\bibfnamefont {T.}~\bibnamefont {Ohta}}, \ and\ \bibinfo {author}
  {\bibfnamefont {H.}~\bibnamefont {L{\"o}wen}},\ }\href@noop {} {\bibfield
  {journal} {\bibinfo  {journal} {The Journal of Chemical Physics}\ }\textbf
  {\bibinfo {volume} {139}},\ \bibinfo {pages} {104906} (\bibinfo {year}
  {2013})}\BibitemShut {NoStop}%
\bibitem [{\citenamefont {Bitzek}\ \emph {et~al.}(2006)\citenamefont {Bitzek},
  \citenamefont {Koskinen}, \citenamefont {G{\"a}hler}, \citenamefont
  {Moseler},\ and\ \citenamefont {Gumbsch}}]{bitzek2006structural}%
  \BibitemOpen
  \bibfield  {author} {\bibinfo {author} {\bibfnamefont {E.}~\bibnamefont
  {Bitzek}}, \bibinfo {author} {\bibfnamefont {P.}~\bibnamefont {Koskinen}},
  \bibinfo {author} {\bibfnamefont {F.}~\bibnamefont {G{\"a}hler}}, \bibinfo
  {author} {\bibfnamefont {M.}~\bibnamefont {Moseler}}, \ and\ \bibinfo
  {author} {\bibfnamefont {P.}~\bibnamefont {Gumbsch}},\ }\href@noop {}
  {\bibfield  {journal} {\bibinfo  {journal} {Physical review letters}\
  }\textbf {\bibinfo {volume} {97}},\ \bibinfo {pages} {170201} (\bibinfo
  {year} {2006})}\BibitemShut {NoStop}%
\bibitem [{\citenamefont {Shimada}\ \emph {et~al.}(2018)\citenamefont
  {Shimada}, \citenamefont {Mizuno}, \citenamefont {Wyart},\ and\ \citenamefont
  {Ikeda}}]{shimada2018spatial}%
  \BibitemOpen
  \bibfield  {author} {\bibinfo {author} {\bibfnamefont {M.}~\bibnamefont
  {Shimada}}, \bibinfo {author} {\bibfnamefont {H.}~\bibnamefont {Mizuno}},
  \bibinfo {author} {\bibfnamefont {M.}~\bibnamefont {Wyart}}, \ and\ \bibinfo
  {author} {\bibfnamefont {A.}~\bibnamefont {Ikeda}},\ }\href@noop {}
  {\bibfield  {journal} {\bibinfo  {journal} {Physical Review E}\ }\textbf
  {\bibinfo {volume} {98}},\ \bibinfo {pages} {060901} (\bibinfo {year}
  {2018})}\BibitemShut {NoStop}%
\bibitem [{\citenamefont {Wyart}(2005)}]{wyart2005rigidity}%
  \BibitemOpen
  \bibfield  {author} {\bibinfo {author} {\bibfnamefont {M.}~\bibnamefont
  {Wyart}},\ }\href@noop {} {\bibfield  {journal} {\bibinfo  {journal} {arXiv
  preprint cond-mat/0512155}\ } (\bibinfo {year} {2005})}\BibitemShut {NoStop}%
\end{thebibliography}%
\end{document}